\documentclass[prd,12pt,a4paper,nofootinbib,showkeys,tightenlines]{revtex4}
\usepackage[left=2.5cm,right=2.5cm,top=2.5cm,bottom=2.5cm]{geometry}

\usepackage[frenchb,english]{babel}
\usepackage[latin1]{inputenc}
\usepackage{enumerate}
\usepackage{amsmath}
\usepackage{amsthm}
\usepackage{amssymb}
\usepackage{graphicx}
\usepackage{floatflt}
\usepackage{fancybox}
\usepackage{appendix}
\usepackage{enumitem}
\usepackage{subfigure}
\usepackage{array}

\usepackage{hyperref}
\usepackage[b]{esvect}

\usepackage{color}

\renewcommand{\Re}{\textrm{Re}}
\renewcommand{\Im}{\textrm{Im}}

\newcommand{\om}{\omega}

\newcommand{\lam}{\lambda}
\newcommand{\Gam}{\Gamma}

\newcommand{\varep}{E} %UNUSUAL

\newcommand{\pal}{\alpha}
\newcommand{\pbet}{\beta}
\newcommand{\pgam}{\gamma}
\newcommand{\pdel}{\delta}

\renewcommand{\vec}[1]{\mathbf{#1}}
\newcommand{\p}{\partial}

\newcommand{\be} {\begin{equation}}
\newcommand{\ee} {\end{equation}}
\newcommand{\bsub}{\begin{subequations}}
\newcommand{\esub}{\end{subequations}}
\newcommand{\bea}{\begin{eqnarray}}
\newcommand{\eea}{\end{eqnarray}}
\newcommand{\bi} {\begin{itemize}}
\newcommand{\ei} {\end{itemize}}
\newcommand{\ben} {\begin{enumerate}}
\newcommand{\een} {\end{enumerate}}
\newcommand{\bmat} {\begin{pmatrix}}
\newcommand{\emat} {\end{pmatrix}}
\newcommand{\bal} {\begin{aligned}}
\newcommand{\eal} {\end{aligned}}
\newcommand{\btab}{\begin{tabular}}
\newcommand{\etab}{\end{tabular}}

\newcommand{\eq}[1]{equation~\eqref{#1}}

\begin{document}
\selectlanguage{english}

\title{Topological two-dimensional Su-Schrieffer-Heeger analogue acoustic networks: 

Total reflection at corners and corner induced modes}

\author{Antonin Coutant}
\email{antonin.coutant@univ-lemans.fr}
\affiliation{Laboratoire d'Acoustique de l'Université du Mans, Unite Mixte de Recherche 6613, Centre National de la Recherche Scientifique, Avenue O. Messiaen, F-72085 LE MANS Cedex 9, France}
\affiliation{Institut de Math\' ematiques de Bourgogne (IMB), UMR 5584, CNRS, Universit\' e de Bourgogne Franche-Comt\' e, F-21000 Dijon, France}

\author{Vassos Achilleos} 
\email{achilleos.vassos@univ-lemans.fr}
\affiliation{Laboratoire d'Acoustique de l'Université du Mans, Unite Mixte de Recherche 6613, Centre National de la Recherche Scientifique, Avenue O. Messiaen, F-72085 LE MANS Cedex 9, France}

\author{Olivier Richoux}
\email{olivier.richoux@univ-lemans.fr}
\affiliation{Laboratoire d'Acoustique de l'Université du Mans, Unite Mixte de Recherche 6613, Centre National de la Recherche Scientifique, Avenue O. Messiaen, F-72085 LE MANS Cedex 9, France}

\author{Georgios Theocharis}
\email{georgios.theocharis@univ-lemans.fr}
\affiliation{Laboratoire d'Acoustique de l'Université du Mans, Unite Mixte de Recherche 6613, Centre National de la Recherche Scientifique, Avenue O. Messiaen, F-72085 LE MANS Cedex 9, France}

\author{Vincent Pagneux}
\email{vincent.pagneux@univ-lemans.fr}
\affiliation{Laboratoire d'Acoustique de l'Université du Mans, Unite Mixte de Recherche 6613, Centre National de la Recherche Scientifique, Avenue O. Messiaen, F-72085 LE MANS Cedex 9, France}

\date{\today}

\begin{abstract}
In this work, we investigate some aspects of an acoustic analogue of the two-dimensional Su-Schrieffer-Heeger model. The system is composed of alternating cross-section tubes connected in a square network, which in the limit of narrow tubes is described by a discrete model coinciding with the two-dimensional Su-Schrieffer-Heeger model. This model is known to host topological edge waves, and we develop a scattering theory to analyze how these waves scatter on edge structure changes. We show that these edge waves undergo a perfect reflection when scattering on a corner, incidentally leading to a new way of constructing corner modes. It is shown that reflection is high for a broad class of edge changes such as steps or defects. We then study consequences of this high reflectivity on finite networks. Globally, it appears that each straight part of edges, separated by corners or defects, hosts localized edge modes isolated from their neighbourhood. 
\end{abstract}


\maketitle

%\tableofcontents

%%%%%%%%%%%%%%%%%%%%%%%%%%%%%%%%%%%%%%%%%%%%%%%%%%%
%%%%%%%%%%%%%%%%%%%%%%%%%%%%%%%%%%%%%%%%%%%%%%%%%%%
%%%%%%%%%%%%%%%%%%%%%%%%%%%%%%%%%%%%%%%%%%%%%%%%%%%
%
%							INTRODUCTION
%
%%%%%%%%%%%%%%%%%%%%%%%%%%%%%%%%%%%%%%%%%%%%%%%%%%%
%%%%%%%%%%%%%%%%%%%%%%%%%%%%%%%%%%%%%%%%%%%%%%%%%%%
%%%%%%%%%%%%%%%%%%%%%%%%%%%%%%%%%%%%%%%%%%%%%%%%%%%
\section{Introduction}

In the last decade, a rapid growth of interest has emerged in the applications of the field of topological insulators to various types of classical waves, as in photonics~\cite{Ozawa19}, mechanics~\cite{Huber16}, or acoustics~\cite{Zhang18,Ma19}. Topological phases were first discovered in the context of the quantum Hall effect (QHE)~\cite{Thouless82} and later, the quantum spin Hall effect (QSHE)~\cite{Kane05}. In these systems, one of the most appealing properties of the topological edge waves is their ability to perfectly transmit through defects. This perfect transmission is ensured by the presence of a single unidirectional edge mode (QHE), or the impossibility of changing the direction of propagation without flipping the spin (QSHE). 

Acoustic waves do not have an intrinsic spin, and mimicking a magnetic field requires to break time reversal. It has been proposed to use a background velocity of air~\cite{Khanikaev15}, or time dependent material properties~\cite{Fleury16} to achieve that, but this comes with challenging difficulties for experiments, such as dissipation, instabilities or noise.  An alternative approach is to rely on crystalline symmetries or hopping modulations to build topologically non-trivial systems~\cite{Fu11,He16,Wu16,Liu17b}. However, transport properties of edge waves in these systems are much less understood. 

In this work, we study the transport properties of edge waves in an acoustic network governed the so-called two-dimensional (2D) Su-Schrieffer-Heeger (SSH) model~\cite{Liu17,Liu18,Obana19,Zheng19}. The network is obtained by connected air channels on a square lattice with changes of cross-section. In the limit of narrow tubes, the acoustic network can be described by a discrete model on a lattice, which coincide with the 2D SSH model. This approach contrasts with the traditional tight binding approximation (TBA), and has the advantages of being valid over a broad frequency range with easily tunable parameters. We then analyze the scattering properties of topological edge waves on an irregular boundary (corner, step or with a defect). We find that, quite surprisingly, edge waves are strongly reflected by all types of irregularity. In fact, we show that reflection is total on a corner (90$°$ turn of the edge), and non-zero transmission across other irregularities only occur by evanescent coupling, and is therefore strongly suppressed. This has interesting consequences in finite-sized networks, where each straight edge acts as a one-dimensional cavity with edge waves forming cavity modes uncoupled to other edges. 

The paper is organized as follows. In section~\ref{Network_Sec} we describe the acoustic network in the narrow limit and obtain the discrete eigenvalue problem coinciding with the 2D SSH model. In section~\ref{SemiInf_Net_Sec}, we describe edge waves on an infinite straight edge, and show that reflection is total on a corner. Then, in section~\ref{Waveguide_Sec}, we develop a scattering theory for edge waves on changing boundaries (steps or defects), and confirm the strong reflection. Finally, in section~\ref{Finite_Net_Sec} we analyze finite networks and cavity modes formed by edge waves.

%%%%%%%%%%%%%%%%%%%%%%%%%%%%%%%%%%%%%%%%%%%%%%%%%%%
%%%%%%%%%%%%%%%%%%%%%%%%%%%%%%%%%%%%%%%%%%%%%%%%%%%
%%%%%%%%%%%%%%%%%%%%%%%%%%%%%%%%%%%%%%%%%%%%%%%%%%%
%
%							2D 		SSH		MODEL
%
%%%%%%%%%%%%%%%%%%%%%%%%%%%%%%%%%%%%%%%%%%%%%%%%%%%
%%%%%%%%%%%%%%%%%%%%%%%%%%%%%%%%%%%%%%%%%%%%%%%%%%%
%%%%%%%%%%%%%%%%%%%%%%%%%%%%%%%%%%%%%%%%%%%%%%%%%%%
\section{From acoustic networks to two-dimensional SSH}
\label{Network_Sec}
\begin{figure}[htp]
\centering
\includegraphics[width=0.5\columnwidth]{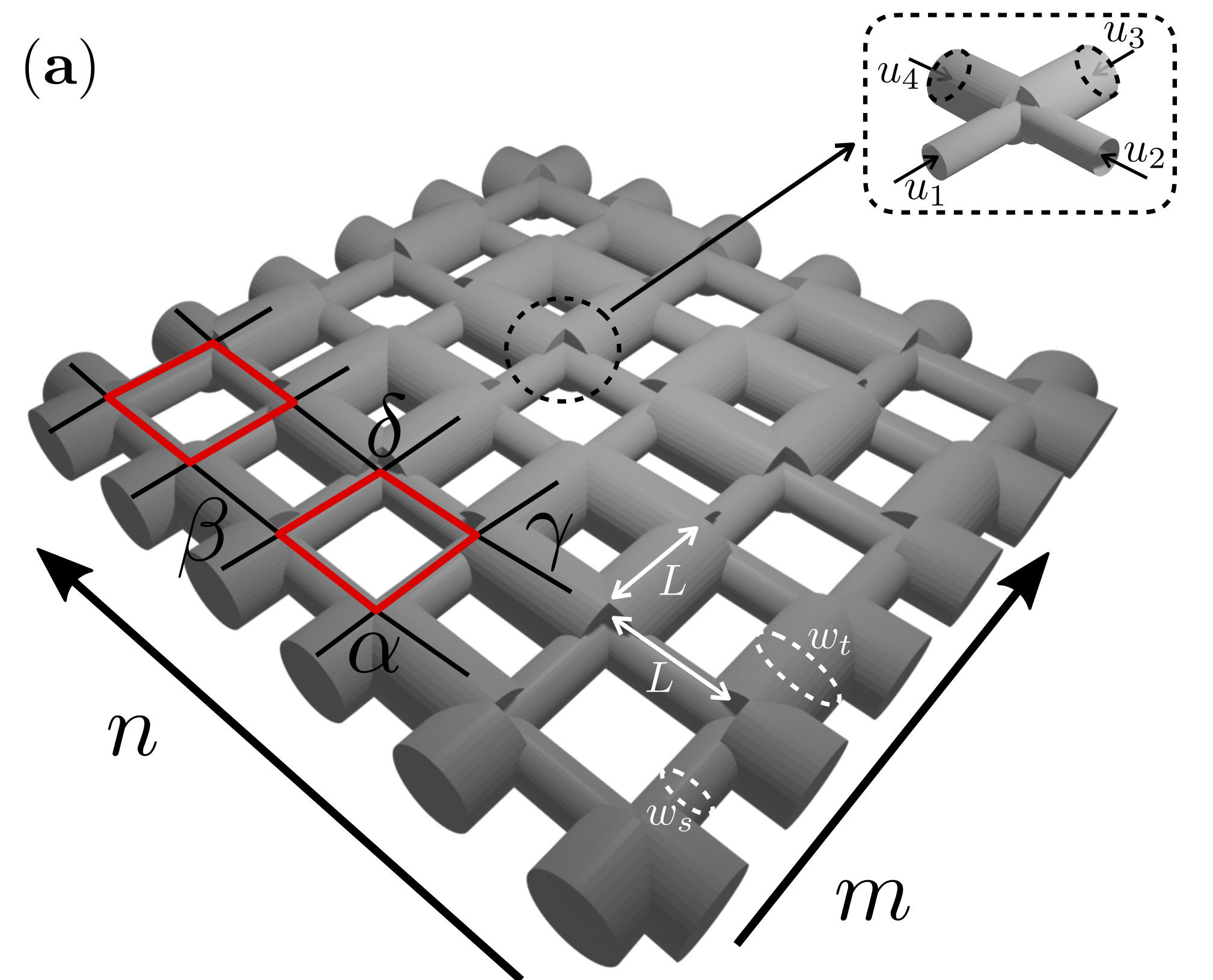} 
\includegraphics[width=0.4\columnwidth]{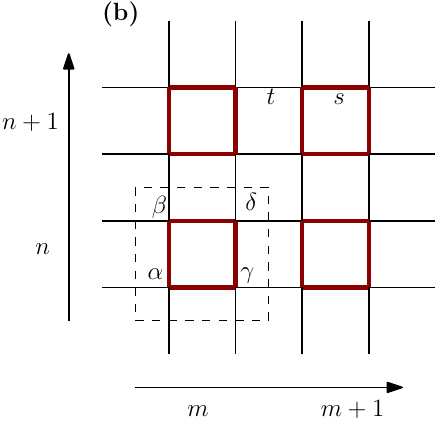} 

\includegraphics[width=0.9\columnwidth]{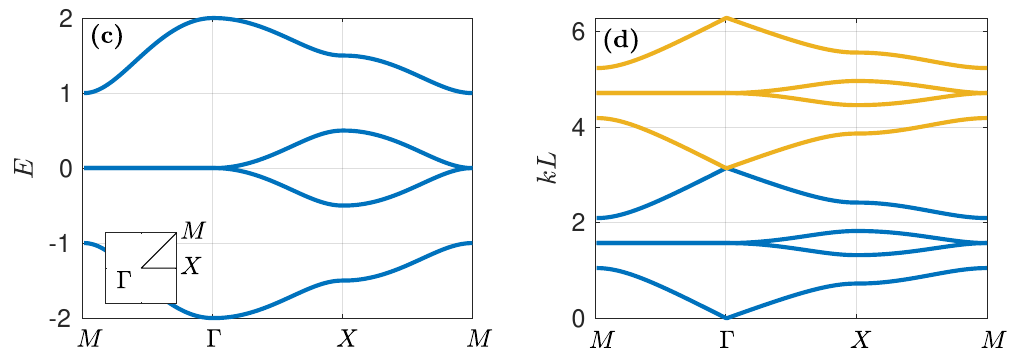}
\medskip

\includegraphics[width=0.75\columnwidth]{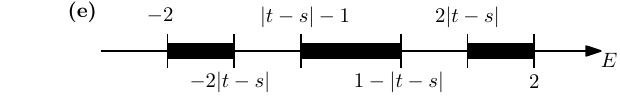}
\caption{(a) Representation of an acoustic network. (b) Schematic representation of the acoustic network as a two-dimensional SSH model. (c) Dispersion relation in energy $\varep$ with $s = 0.25$ and $t=0.75$. (d) Corresponding dispersion relation in dimensionless frequency $kL$ ($s = 0.25$ and $t=0.75$). (e) Spectrum: allowed energy ranges (bands) are shown in bold and depend on the single parameter $|t-s|$. 
}
\label{2D_SSH_Fig} 
\end{figure}

In this work, we consider sound waves propagating in air channels connected in a periodic square network. Each unit cells consist in four intersections mutually connected by channels of section $w_s$, and connected to neighbouring cells by channels of section $w_t$ (Fig.~\ref{2D_SSH_Fig}-(a)). Our treatment is based on the fact that if the length of the tubes $L$ is much larger that the transverse lengths, $L^2 \gg w$ (which is short for both $L^2 \gg w_s$, and  $L^2 \gg w_t$), the system can be accurately described by a discrete model~\cite{Depollier90,Zheng19,Zheng20}. 

To see this, consider an acoustic wave of fixed frequency $\om = k c_0$, with $k$ the wavenumber and $c_0$ the speed of sound. The acoustic field is described by a harmonic pressure field $p(x,y,z) e^{-i \om t}$. To solve the problem of harmonic waves in the network, we claim that it is sufficient to know the value of the pressure at each intersection, and that this set of pressure values obey a discrete problem. In the limit of narrow tubes, i.e. $L^2 \gg w$, the propagation inside each tube is one-dimensional (monomode propagation), which means the Helmholtz equation reduces to $p'' + k^2 p=0$ (where derivative is with respect to the cartesian coordinate along the considered tube). Moreover, at each intersection, pressure is continuous and the sum of flow rates vanishes.

Let us consider one such intersection. Knowing the pressure $p$ and acoustic velocities $u_j$ at an intersection (see Fig.~\ref{2D_SSH_Fig}-(a)), we can integrate the one-dimensional Helmholtz equation to obtain the pressure $p_j$ at a neighbouring intersection:
\be \label{1D_Helm_eq}
\cos(kL) p + i \sin(kL) \rho_0 c_0 u_j = p_j, 
\ee
where $\rho_0$ is air density, $i^2=-1$, $j=1..4$, and $u_j$ the acoustic velocity in the tube linking $p$ and $p_j$ and evaluated at the entrance of the $p$ intersection (see Fig.~\ref{2D_SSH_Fig}-(a)). We can now write the conservation of the acoustic flow rate: 
\be
w_s u_1 + w_s u_2 + w_t u_3 + w_t u_4 = 0,  
\ee
and use it to eliminate the acoustic velocities from \eq{1D_Helm_eq}. Summing \eq{1D_Helm_eq} over $j$, and using debit conservation leads to an equation only on pressure: 
\be \label{Discrete_pressure_eq}
(w_s+w_t) \varep p = w_s p_1 + w_s p_2 + w_t p_3 + w_t p_4,  
\ee
where 
\be
\varep=2\cos(kL). 
\ee
We now derive this equation for every intersection. Since each unit cell has four intersections (see Fig.~\ref{2D_SSH_Fig}-(b)), we label the pressure according to the cell coordinate $(m,n)$, and the position inside the cell: $\alpha_{m,n}$ in the lower left, $\beta_{m,n}$ in the upper left, $\gamma_{m,n}$ in the lower right, and $\delta_{m,n}$ in the upper right (see Fig.~\ref{2D_SSH_Fig}-(b)). Hence, the discrete \eq{Discrete_pressure_eq} applied to each intersection leads to the system 
\bsub \label{2D_SSH_master_Eq} \bea
s \pbet_{m,n} + t \pbet_{m,n-1} + s \pgam_{m,n} + t \pgam_{m-1,n} &=& \varep \pal_{m,n} , \label{2DSSH_alpha} \\
s \pal_{m,n} + t \pal_{m,n+1} + s \pdel_{m,n} + t \pdel_{m-1,n} &=& \varep \pbet_{m,n} , \label{2DSSH_beta} \\
s \pal_{m,n} + t \pal_{m+1,n}  + s \pdel_{m,n} + t \pdel_{m,n-1} &=& \varep \pgam_{m,n} , \label{2DSSH_gamma} \\
s \pbet_{m,n} + t \pbet_{m+1,n} + s \pgam_{m,n} + t \pgam_{m,n+1} &=& \varep \pdel_{m,n} , \label{2DSSH_delta} 
\eea \esub
with the effective coupling coefficients $s=w_s/(w_s+w_t)$ and $t=w_t/(w_s+w_t)$. Solving this system gives eigenvalue $\varep$, which in turn gives us several allowed frequencies $kL=\pm \mathrm{acos}(\varep/2)+2\ell \pi$ ($\ell \in \mathbb Z)$. The complete pressure field can then be obtained by solving the boundary value problem $p'' + k^2 p=0$ with known pressure on both ends. This boundary value problem has a unique solution except on the Bragg frequencies $kL = \ell \pi$, which means the discrete system of \eq{2D_SSH_master_Eq} may miss these solutions, that must be added by hand. This is a minor concern for us though, since we are not interested in the vicinity of these frequencies. 

We now underline that \eq{2D_SSH_master_Eq} has the form of a stationary Schrödinger equation on a lattice $H \cdot X = \varep X$, with the vector $X$ gathering all pressure values $\pal_{m,n}$, $\pbet_{m,n}$, $\pgam_{m,n}$ and $\pdel_{m,n}$. The Hamiltonian $H$ is that of the 2D SSH model~\cite{Liu17,Liu18,Obana19}, and contains only the geometrical parameters $s$ and $t$ as hopping coefficients. By analogy to quantum systems, we refer to $\varep$ as the energy.

Lattice Hamiltonian are traditionally obtained through the TBA~\cite{Ashcroft}, or couple mode theory in the context of classical waves~\cite{Haus91,Ozawa19}. It is however worth emphasizing the main advantages of the acoustic network approach. First, it is valid on a broad range of frequencies, since it only requires $L^2\gg w$ and monomode propagation, unlike TBA which focuses on the vicinity of a typical frequency of resonators (or energy levels) being coupled. Second, coupling constants are given by ratio of cross-sections, and therefore rather easy to control experimentally. This contrasts with the TBA approach where coupling constant are given in terms of wave functions overlaps, and hence usually directly fitted from numerical simulations or experiment. This acoustic network approach is very similar to that of transmission line networks in photonics~\cite{Zhang98,Cheung04,Jiang19} or quantum graphs~\cite{Kuchment07,Kuchment08}. 

\subsection{Two-dimensional SSH model and dispersion relation}
\label{Bloch_Sec}

The set of equation \eqref{2D_SSH_master_Eq} gives an eigenvalue problem for $\varep$, which coincide with the so-called two-dimensional SSH model~\cite{Liu17,Liu18,Obana19,Zheng19}. It turns out that this model is separable~\cite{Benalcazar20,Zhu20,Cerjan20,Coutant20}, similarly to the Helmholtz equation in a rectangular domain. Hence, we look for solutions of the form 
\be \label{Full_2D_TensorProd}
\bmat \pal_{m,n} \\ \pbet_{m,n} \\ \pgam_{m,n} \\ \pdel_{m,n} \emat = \bmat \psi_A^m \varphi_A^n \\ \psi_A^m \varphi_B^n \\ \psi_B^m \varphi_A^n \\ \psi_B^m \varphi_B^n \emat . 
\ee
Now, the left hand-side is a solution of the 2D problem of \eq{2D_SSH_master_Eq} if 
\bsub \label{Long_SSH} \bea
\varep_x \psi_A^m &=& t \psi_B^{m-1} + s \psi_B^{m} , \\ 
\varep_x \psi_B^m &=& t \psi_A^{m+1} + s \psi_A^{m} ,  
\eea \esub
and 
\bsub \label{Transverse_SSH} \bea
\varep_y \varphi_A^n &=& t \varphi_B^{n-1} + s \varphi_B^{n} , \\ 
\varep_y \varphi_B^n &=& t \varphi_A^{n+1} + s \varphi_A^{n} , 
\eea \esub
with $\varep = \varep_x + \varep_y$. This is explained in more details in appendix~\ref{Sep_App}. In an infinite network, we can look for Bloch wave solution: 
\be
\bmat \pal_{m,n} \\ \pbet_{m,n} \\ \pgam_{m,n} \\ \pdel_{m,n} \emat = e^{i m q_x + i n q_y} \bmat \pal \\ \pbet \\ \pgam \\ \pdel \emat. 
\ee
Moreover, using separability, the dispersion relation can be written
\be \label{2D_DispRel}
\varep(q_x,q_y) = \varep_{1D}(q_x) + \varep_{1D}(q_y), 
\ee 
with $\varep_{1D}$ the dispersion relation of the one-dimensional problem. This is obtained by solving equations~\eqref{Long_SSH} and \eqref{Transverse_SSH} with one-dimensional Bloch waves: 
\be \label{1D_SSH_Bloch}
\bmat 0 & s + t e^{-iq} \\ s + t e^{iq} & 0 \emat \cdot \Phi = \varep_{1D}(q) \Phi, 
\ee
with $\Phi = \bmat \psi_A & \psi_B \emat^T$ or $\Phi = \bmat \varphi_A & \varphi_B \emat^T$. This leads to the dispersion relation 
\be \label{1D_DispRel}
\varep_{1D} = \pm |s+t e^{iq}| . 
\ee
The full dispersion relation \eqref{2D_DispRel} is shown in Fig.~\ref{2D_SSH_Fig}-(c). The 1D dispersion relation \eqref{1D_DispRel} consists in two branches of opposite energies. Combining two of them through equation \eqref{2D_DispRel}, we see that the 2D dispersion relation has four bands. Noticing that $|s+te^{iq}|$ varies from $|t-s|$ to $1$, the lowest bands varies from $-2$ to $-2|t-s|$, two middle bands coincide and are comprised between $|t-s|-1$ and $1-|t-s|$, and the upper varies from $2|t-s|$ to $2$. This is illustrated in Fig.~\ref{2D_SSH_Fig}(e), in particular, when $|t-s| > 1/3$, i.e. $2|t-s|>1-|t-s|$, the energy spectrum has two full bandgaps, as illustrated in Fig.~\ref{2D_SSH_Fig}-(e). However, we point out that although this model has four bands for the energy $\varep$, because of the relation $\varep=2\cos(kL)$, it corresponds of course to an infinite number of bands for the reduced frequency $kL$ (compare Fig.~\ref{2D_SSH_Fig}-(c) and (d), where different colors show different frequency solutions associated with the same energy $\varep$). 

%%%%%%%%%%%%%%%%%%%%%%%%%%%%%%%%%%%%%%%%%%%%%%%%%%%
%%%%%%%%%%%%%%%%%%%%%%%%%%%%%%%%%%%%%%%%%%%%%%%%%%%
%%%%%%%%%%%%%%%%%%%%%%%%%%%%%%%%%%%%%%%%%%%%%%%%%%%
%
%						2D semi-infinite network
%
%%%%%%%%%%%%%%%%%%%%%%%%%%%%%%%%%%%%%%%%%%%%%%%%%%%
%%%%%%%%%%%%%%%%%%%%%%%%%%%%%%%%%%%%%%%%%%%%%%%%%%%
%%%%%%%%%%%%%%%%%%%%%%%%%%%%%%%%%%%%%%%%%%%%%%%%%%%
\section{Edge waves and corner effects}
\label{SemiInf_Net_Sec}
In this section, we consider a network with a horizontal edge. This is obtained by adding together unit cells for $m \in \mathbb Z$ and $n=1..\infty$. Moreover, at the lower end of the network, we add open ended channels, where the acoustic pressure vanishes. Hence, we must now solve the eigenvalue problem of \eq{2D_SSH_master_Eq} with the additional boundary conditions $\pbet_{m,n=0} = \pdel_{m,n=0} = 0$. If the system has nontrivial topological properties~\cite{Hasan10,Xiao14,Ma19}, edge waves appear and propagate along the boundary while being evanescent inside the network. This was shown explicitly in the 2D SSH model: edge waves are present whenever $s<t$, which correspond to a nonzero (quantized) 2D Zak phase~\cite{Liu17,Liu18,Obana19}. Since our acoustic network is exactly described by the 2D SSH model through equation~\eqref{2D_SSH_master_Eq}, we conclude that edge waves will be present whenever $s<t$. From now on, we assume that this condition holds. The rest of this paper is devoted to the study of the scattering properties of these edge waves.

\subsection{Infinite horizontal edge}
\label{SemInf_OneEdge_Sec}
\begin{figure}[htp]
\centering
\includegraphics[width=0.7\columnwidth]{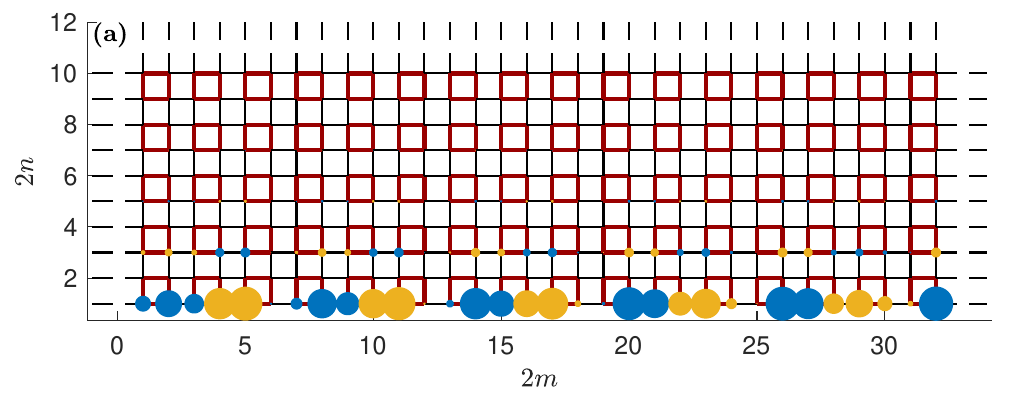}
\includegraphics[width=0.29\columnwidth]{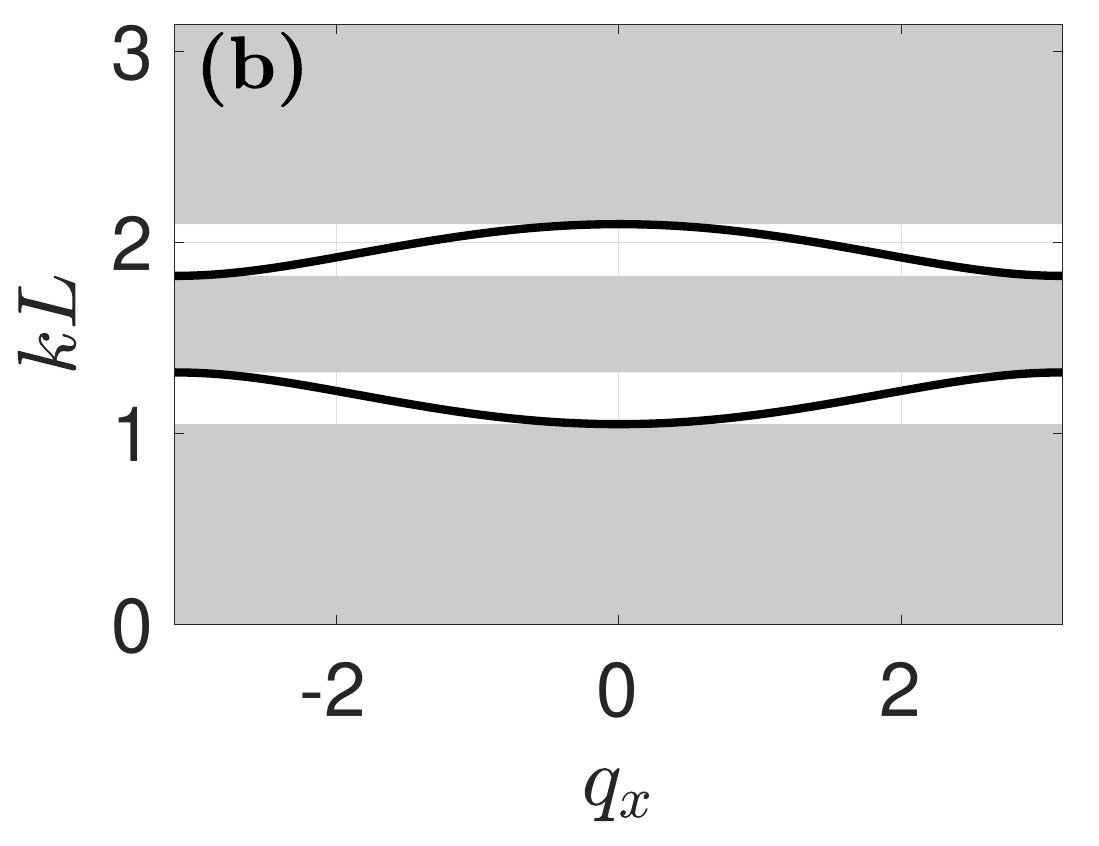}
\caption{(a) Pressure amplitude $\Re(P_m)$ (arbitrary units) of edge modes with $\varep=0.7$, $s=0.25$ and $t=0.75$. Positive (resp. negative) values of the pressure are shown as yellow (resp. blue) circles at each intersections. Note that the network extends periodically in the upper, left and right directions. (b) Dispersion relation of the edge mode: $kL$ as a function of $q_x$ (with $E = 2\cos(kL)$). Grey area show the frequency ranges of bulk waves (pass band of Fig.~\ref{2D_SSH_Fig}-(d)).}
\label{WG_EdgeModes_Fig} 
\end{figure}

Using the separability of the eigenvalue problem \eqref{2D_SSH_master_Eq} we can obtain a closed-form expression for edge waves along this boundary. For this, we look for solutions with separation of variables, i.e. of the form of~\eq{Full_2D_TensorProd}. The horizontal part $\psi_{A/B}^m$ is a Bloch wave solution of \eq{Long_SSH}, and the vertical part $\varphi_{A/B}^n$ is an evanescent mode solution of \eq{Transverse_SSH}, such that the two-dimensional mode satisfies the boundary conditions $\pbet_{m,n=0} = \pdel_{m,n=0} = 0$ (Fig.~\ref{WG_EdgeModes_Fig}-(a)). This lead to a solution 
\be \label{2D_sinf_EW}
\bmat \pal_{m,n} \\ \pbet_{m,n} \\ \pgam_{m,n} \\ \pdel_{m,n} \emat = e^{i m q_x} \bmat e^{-i \theta_x} \\ 0 \\ 1 \\ 0 \emat (-s/t)^n. 
\ee
with 
\be \label{Bloch_Phase}
e^{i \theta_x} = \frac{s+t e^{iq_x}}{|s+t e^{iq_x}|} . 
\ee
Equation~\eqref{Bloch_Phase} comes from solving the horizontal Bloch problem \eqref{Long_SSH}. Since $\varphi_B^n = 0$ from \eq{2D_sinf_EW}, the transverse \eq{Transverse_SSH} gives $\varep_y = 0$, and hence, the real Bloch wavenumber $q_x$ satisfies the dispersion relation 
\be \label{EdgeWave_DispRel}
\varep = |s + t e^{iq_x}|, 
\ee
which is nothing else than the positive branch of the 1D dispersion relation of equation \eqref{1D_DispRel}. Of course, equation~\eqref{2D_sinf_EW} also manifestly satisfies the boundary conditions $\pbet_{m,n=0} = \pdel_{m,n=0} = 0$. If we fix the energy $\varep$ to be $|s-t| < \varep < 1$, we see from equations~\eqref{1D_DispRel} and \eqref{EdgeWave_DispRel} that the Bloch wavenumber $q_x$ is real, hence \eq{2D_sinf_EW} describes a propagating wave. Moreover, we also see from \eq{2D_sinf_EW} that this waves is localized on the edge by construction: its amplitude decreases to 0 for increasing $n$ (we recall that we assumed $s<t$). This edge wave solution is shown in Fig.~\ref{WG_EdgeModes_Fig}. Notice that at fixed energy $\varep$, there are two opposite solutions for $q_x$, corresponding to a left and a right moving wave. There are also edge waves in the interval $-1 < E < -|t-s|$, and their properties are symmetric with respect to \eq{2D_sinf_EW}, see Fig.~\ref{WG_EdgeModes_Fig}. Indeed, edge waves for negative energies can be obtained by flipping the sign of the $\beta$ and $\gamma$ amplitudes (see appendix~\ref{Chiral_App}). To simplify the discussion, from now on we only consider $E>0$.

\subsection{Total reflection on an isolated corner}
\label{Corner_Scatt_Sec}
\begin{figure}[htp]
\centering
\includegraphics[width=0.9\columnwidth]{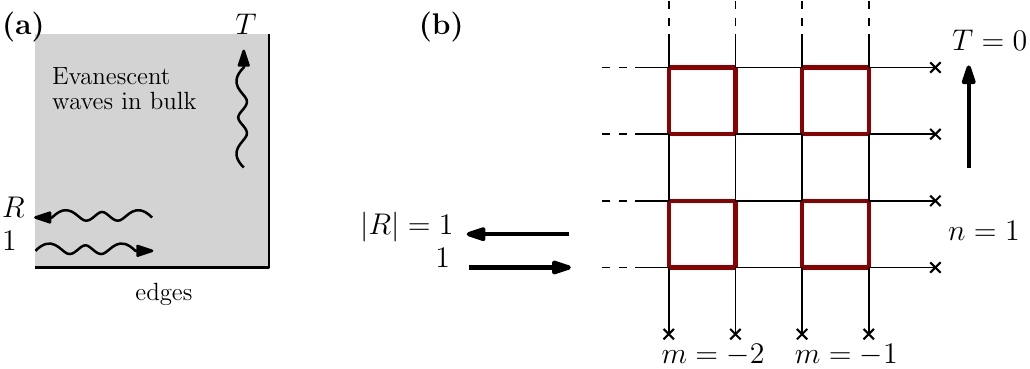}
\caption{(a) Illustration of the scattering of edge waves by a corner, in a range of frequency where no bulk wave can propagate. (b) Structure of a corner in a 2D SSH network, semi-infinite in both directions. Crosses mark open ends where acoustic pressure vanishes. Arrows schematically represent an incoming edge wave from the left, totally reflected on the corner.}
\label{WG_Corner_Fig} 
\end{figure}

We now investigate how an edge wave as in \eq{2D_sinf_EW} scatters on an isolated corner, as illustrated in Fig.~\ref{WG_Corner_Fig}. For this, we simply assume a vertical edge of the network along $m=0$, i.e. $m=-\infty .. -1$. Similarly, open end channels are added to the vertical edge so that solutions satisfy the extra boundary conditions $\pal_{m=0, n} = \pbet_{m=0, n} = 0$. In this configuration, edge waves can propagate on both the horizontal and vertical edges. The question is whether a wave localized on one edge incoming on the corner is converted to an outgoing edge wave on the other edge. Additionally, we assume that in the range of energies such that edge waves are propagative ($|t-s| < E < 1$, see section~\ref{SemInf_OneEdge_Sec}) lies inside the bulk gap ($1-|t-s| < E < 2|t-s|$, see Fig.~\ref{2D_SSH_Fig}-(e)). 

We focus on the case of a horizontal edge wave incoming from the left on the corner. Due to separability, it is possible to build the solution of the problem as a superposition of two edge waves with $\pm q_x$. Indeed, from \eqref{2D_sinf_EW}, we see that the solution satisfying all boundary conditions is 
\be \label{Corner_Scatt_eq}
\bmat \pal_{m,n} \\ \pbet_{m,n} \\ \pgam_{m,n} \\ \pdel_{m,n} \emat = e^{i m q_x} \bmat e^{-i \theta_x} \\ 0 \\ 1 \\ 0 \emat (-s/t)^n - e^{-i m q_x - 2 i \theta_x} \bmat e^{i \theta_x} \\ 0 \\ 1 \\ 0 \emat (-s/t)^n, 
\ee
where we assume $q_x > 0$. Since this solution does not involve an edge mode on the edge along the $y$-axis, this means that the transmission to the other edge vanishes and the reflection coefficient is $R=-e^{-2i \theta_x}$, hence $|R|=1$. Notice that even if we had considered an energy $\varep$ with propagating bulk waves (see Fig.~\ref{WG_DispRel_Fig}-(b,c)), \eq{Corner_Scatt_eq} would still be valid, meaning that edge waves do not scatter in the bulk when hitting an isolated corner.

\subsection{Corner mode from ``edge-corner correspondence''}

It is also interesting to notice the non-trivial phase of the reflection coefficient. In fact, this phase is a marker of the topological structure of the edge wave band~\cite{Xiao14}, which means the higher order topology of the network~\cite{Xie18,Ota19,Zhu20,Xu20,Coutant20}. Indeed, in the one-dimensional SSH model, it was shown that the relative phase between the two components of Bloch waves encodes the presence or absence of an edge mode~\cite{Delplace11,Dalibard18}, leading to a constructive proof of the bulk-edge correspondence in one dimension. Interestingly, the same argument can be employed here, in two dimensions, to explain the presence of corner modes from the phase of the reflection coefficient. By similarity to the bulk-boundary correspondence, we call this ``edge-corner correspondence''. 

\begin{figure}[htp]
\centering
\includegraphics[width=\columnwidth]{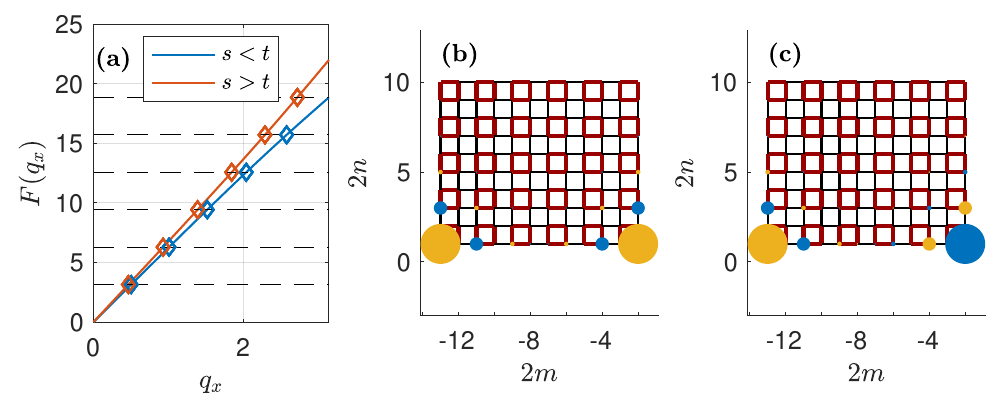}
\caption{(a) We plotted $F(q_x) = (N_x+1)q_x - \theta_x$ such that \eq{Cavity_modes_eq} reads $\sin(F(q_x))=0$. Dashed lines are multiples of $\pi$, and each intersection with $F(q_x)$ corresponds to a cavity mode solution. We shown two cases: $s=0.25$ and $t=0.75$ (blue) and $s=0.75$ and $t=0.25$ (red), both with $N_x=6$. (b,c) Pressure amplitude (arbitrary units) for the corner modes in the topological case ($s=0.25$ and $t=0.75$). Positive and negative values are shown in yellow and blue.}
\label{Corner_Phase_Fig} 
\end{figure}

To understand this, we first point out that although the phase of the reflection coefficient is not an observable in itself, as it depends on a choice of mode basis, it determines the set of cavity modes if the system is closed by a second vertical edge on the left. Assuming the left and right vertical edges are separated by $N_x$ unit cells (Fig.~\ref{Corner_Phase_Fig}-(b,c)), the edge modes become a discrete set of cavity modes. To obtain them, we can start from equation~\eqref{Corner_Scatt_eq} with an undetermined $q_x$. It already satisfies the boundary condition on the left corner, hence we only need to apply the boundary condition to the other corner, namely $\pgam_{m=-(N_x+1), n} = \pdel_{m=-(N_x+1), n} = 0$. This leads to an equation to solve for $q_x$: 
\be \label{Cavity_modes_eq}
\sin((N_x+1)q_x - \theta_x(q_x)) = 0. 
\ee
For each value of $0 < q_x < \pi$ satisfying this equation, there are two modes given by the two branches of the dispersion relation \eqref{1D_DispRel}. The inequalities are strict for the range of $q_x$, since $q_x =0$ or $q_x=\pi$ would lead to a vanishing solution. It is also restricted to $q_x>0$ because $-q_x$ would lead to the same cavity mode. This \eq{Cavity_modes_eq} involves the same phase $\theta_x$ as in the reflection coefficient. Now, when $q_x$ goes from 0 to $\pi$, the phase $\theta_x$ can either go from $0$ to $\pi$ (if $s<t$), or from $0$ back to $0$ (if $s>t$). Hence, there is one extra solution of \eq{Cavity_modes_eq} for $q_x$ in the second case with respect to the first case, which means two extra cavity modes. The missing two modes in the first case correspond to two corner modes. This is illustrated in Fig.~\ref{Corner_Phase_Fig}. Notice that the two corner modes shown in Fig.~\ref{Corner_Phase_Fig}-(b,c) are bound states in the continuum~\cite{Chen19,Benalcazar20,Cerjan20,Coutant20}, since their energy eigenvalue is close to zero in the bulk pass band ($kL$ close to $\pi/2$ in Fig.~\ref{2D_SSH_Fig}-(d)). 

%%%%%%%%%%%%%%%%%%%%%%%%%%%%%%%%%%%%%%%%%%%%%%%%%%%
%%%%%%%%%%%%%%%%%%%%%%%%%%%%%%%%%%%%%%%%%%%%%%%%%%%
%%%%%%%%%%%%%%%%%%%%%%%%%%%%%%%%%%%%%%%%%%%%%%%%%%%
%
%					2D 		SSH		WAVEGUIDES
%
%%%%%%%%%%%%%%%%%%%%%%%%%%%%%%%%%%%%%%%%%%%%%%%%%%%
%%%%%%%%%%%%%%%%%%%%%%%%%%%%%%%%%%%%%%%%%%%%%%%%%%%
%%%%%%%%%%%%%%%%%%%%%%%%%%%%%%%%%%%%%%%%%%%%%%%%%%%
\section{Scattering of edge waves on finite steps}
\label{Waveguide_Sec}
\begin{figure}[htp]
\centering
\includegraphics[width=0.6\columnwidth]{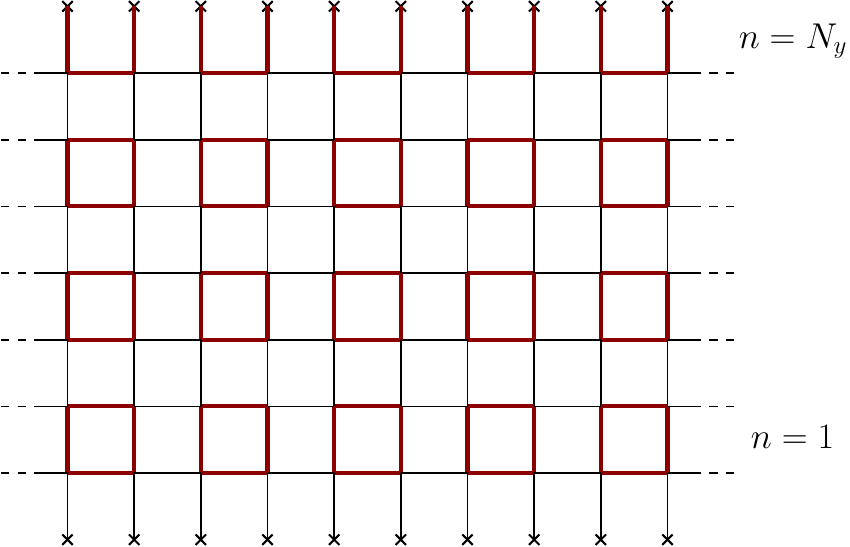}
\caption{Network configuration with finite width $N_y$. Crosses mark open ends where acoustic pressure vanishes. Dashed lines indicate that the configuration repeats itself indefinitely in the $x$-direction.}
\label{WG_Schema_Fig} 
\end{figure}

We now want to analyze how edge waves scatter on a finite sized step, that is, a jump between two parallel edges at different heights. For this we use a multimode expansion for networks with horizontal edges of finite but large width. The employed method is then very similar to multimode methods to describe scattering in acoustic waveguides of varying cross-section~\cite{Pagneux96,Amir97}. The lower edge of the network at $n=1$ is the same as before. On the other side, for $n=N_y$, the unit cells are made of two lower intersections ($\alpha$, $\gamma$) connected vertically with open ended channels, as illustrated in Fig.~\ref{WG_Schema_Fig}. The main advantage of building the upper boundary this way, is that the edge wave of \eq{2D_sinf_EW} is an exact solution of the finite width network waveguide problem. This guarantees that in the relevant energy range ($|t-s| < \varep < 1$, see section~\ref{SemInf_OneEdge_Sec}) there is only one edge wave at the bottom. Of course, for large $N_y$, our results becomes independent of the precise structure of the upper boundary. For different upper edges though, one could have edge waves on the upper edge leading to beating effects with the lower edge wave, which we avoid in our configuration. 

%%%%%%%%%%%%%%%%%%%%%%%%%%%%%%%%%%%%%%%%%%%%%%%%%%%
%							eigenmodeS
%%%%%%%%%%%%%%%%%%%%%%%%%%%%%%%%%%%%%%%%%%%%%%%%%%%
\subsection{Waveguide eigenmodes}
\label{Modes_Sec}
To start, we consider a uniform network waveguide of finite width $N_y$, and we look for modes at a fixed value of energy $\varep$. 
Since the problem is invariant under discrete translations along $x$, we can look for Bloch wave solutions along that direction. 
Using the Bloch wave form on \eq{Full_2D_TensorProd}, we obtain waveguide eigenmodes of the form 
\be \label{2DWG_mode}
P_m = e^{i m q} \bmat \psi_A \vec{\Phi} \\ \psi_B \vec{\Phi} \emat , 
\ee 
where $\vec{\Phi}$ is the vector containing the transverse components $\varphi_A^n$ and $\varphi_B^n$ for $n=1..N_y$. Hence, $\vec{\Phi}$ is a $2N_y-1$ vector. It is a solution of the transverse problem along $y$, obtained by taking \eq{Transverse_SSH} with the boundary conditions $\varphi_B^0 = \varphi_A^{N_y} = 0$. This gives $2N_y-1$ eigenmodes $(\vec{\Phi}^j)_{j=1..2N_y-1}$ associated with the eigenvalues $(\varep_j)_{j=1..2N_y-1}$. Notice that there are $2N_y-1$ of these transverse modes $\vec{\Phi}_j$, which correspond to the number of tubes along the $y$-axis. 
 
\begin{figure}[htp]
\centering
\includegraphics[width=\columnwidth]{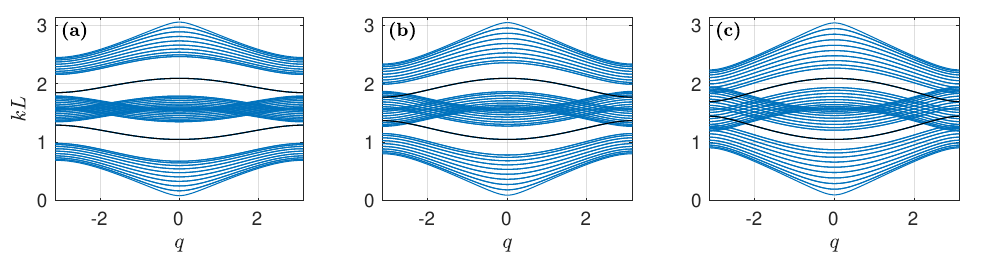}
\caption{Dispersion relation $kL$ as a function of $q$, for a finite width network with $N_y=10$, which means that we have $42$ modes. Black lines show the edge modes. (a): $|t-s| = 0.55$. In this case and for $1/2 < |t-s| < 1$ the edge waves energy range is fully inside the gap. (b): $|t-s| = 0.4$. For $1/3 < |t-s| < 1/2$ the edge waves energy range overlaps with both passing bands and the gap. (c) $|t-s|=0.25$. For $0 < |t-s| < 1/3$ the gap is closed and edge waves have energies inside passing bands.}
\label{WG_DispRel_Fig} 
\end{figure}

Since we look for all solutions at fixed energy $\varep$, we allow $q$ to be complex in \eq{2DWG_mode}.  Indeed, as in scattering problems in waveguides, we must consider all possible evanescent modes, which are excited at section changes~\cite{Pagneux96}. At this level we also define the Bloch eigenvalue $\lam = e^{iq}$, which is complex and characterize the propagation of the mode. Notice that from now on, the wavenumber $q$ is always along the $x$ axis, and hence we dropped the corresponding index. For a given transverse mode $\vec{\Phi}^j$, the components $\bmat \psi_A^j & \psi_B^j \emat^T$ are solutions of the 1D Bloch problem of \eq{1D_SSH_Bloch} with the energy  
\be
 \varep_x = \varep - \varep_j. 
 \ee
As in section~\ref{Bloch_Sec}, we obtain the wavenumber $q_j(\varep)$ with the dispersion relation  
\be
\cos(q_j(\varep)) = \frac{(\varep-\varep_j)^2 -s^2 -t^2}{2st}. 
\ee
Again, we emphasize that unlike section~\ref{Bloch_Sec} where the Bloch wavenumber are chosen real and the energy values follow, here the energy value $\varep$ is given \emph{a priori}, and one must include all solutions for $q_j$, both real and complex. Solutions with real $q_j$ correspond to propagating modes, while non-real $q_j$ are evanescent. Moreover, since there are two opposite solutions for $q_j$ at fixed $\varep$, the total number of modes is twice the number of vertical tubes. In Fig.~\ref{WG_DispRel_Fig} we show the dispersion relation of propagating modes only ($q_j \in \mathbb R$). Indeed, at fixed $q$ there are $4N_y-2$ energy branches, but at fixed energy $\varep$, only some of the $4N_y-2$ solutions correspond to real valued $q$, the other solutions being evanescent. 
In general, evanescent modes must be included in the set of solutions because they can be excited at changes of the edge structure and affect the scattering. 

We now classify the modes of the form \eqref{2DWG_mode} into left going and right going using the Bloch eigenvalue $\lam = e^{iq}$. For this, we split the case of evanescent and propagating waves. If $|\lam|<1$ (resp. $|\lam|>1$) it is evanescent and moves to the right (resp. left). If $|\lam| = 1$, it is propagative, and the direction of propagation is given by the sign of the group velocity: 
\be \label{2D_vg_eq}
v_g^j(\varep) = \frac{\p \varep}{\p q} = - \frac{st}{\varep - \varep_j} \sin(q_j).  
\ee
Alternatively, one can add a small positive imaginary part $\om \to \om - i \nu$ and apply the evanescent wave criterion. In the studied system, both criterions can be shown to be equivalent, but notice that this is not the case for systems with mean flow~\footnote{Inequivalence can typically arise in the presence of mean flows, leading to the phenomenon of convective instabilities, see e.g.~\cite{Crighton91,Coutant19}.}. Using this, a general solution at fixed $\varep$ in the network waveguide reads 
\be \label{Mode_for_Scatt}
P_m = \sum_{j=1}^{2N_y-1}  a_j \lam_{+j}^m \bmat \psi_A^{+j} \vec{\Phi}_j \\ \psi_B^{+j} \vec{\Phi}_j \emat + b_j \lam_{-j}^m \bmat \psi_A^{-j} \vec{\Phi}_j \\ \psi_B^{-j} \vec{\Phi}_j \emat , 
\ee
where $a_j$ and $b_j$ are complex coefficients, and the $+$ indices (resp. $-$ indices) indicate a right (resp. left) going mode.

%%%%%%%%%%%%%%%%%%%%%%%%%%%%%%%%%%%%%%%%%%%%%%%%%%%
%%%%%%%%%%%%%%%%%%%%%%%%%%%%%%%%%%%%%%%%%%%%%%%%%%%
%%%%%%%%%%%%%%%%%%%%%%%%%%%%%%%%%%%%%%%%%%%%%%%%%%%
%
%					     WAVEGUIDE	SCATTERING
%
%%%%%%%%%%%%%%%%%%%%%%%%%%%%%%%%%%%%%%%%%%%%%%%%%%%
%%%%%%%%%%%%%%%%%%%%%%%%%%%%%%%%%%%%%%%%%%%%%%%%%%%
%%%%%%%%%%%%%%%%%%%%%%%%%%%%%%%%%%%%%%%%%%%%%%%%%%%
\subsection{Edge wave scattering}
\label{Step_Scatt_Sec}
%%%%%%%%%%%%%%%%%%%%%%%%%%%%%%%%%%%%%%%%%%%%%%%%%%%
%						2D SCATTERING ON A STEP
%%%%%%%%%%%%%%%%%%%%%%%%%%%%%%%%%%%%%%%%%%%%%%%%%%%

\begin{figure}[htp]
\centering
\includegraphics[width=0.6\columnwidth]{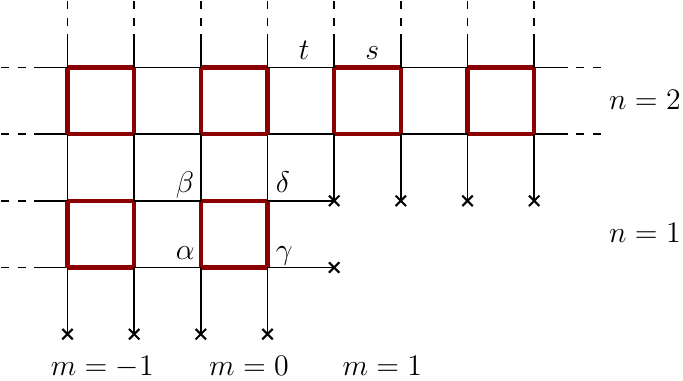}
\caption{Step of $N_0=1$ cells within a waveguide.}
\label{WG_Step_Fig} 
\end{figure}

We now consider a step of height $N_0$ by joining a waveguide of with $N_y$ on the left ($m<0$) with a waveguide of width $N_y-N_0$ on the right ($m \geqslant 0$), as illustrated in Fig.~\ref{WG_Step_Fig} for $N_0=1$. We want to analyze how an edge wave scatter on such a step. This problem is rather similar to the one of a duct wave scattering on a discontinuous change of cross-section, and hence, we shall develop a similar formalism~\cite{Pagneux96,Amir97,Rienstra}, based on mode matching methods. 

\subsubsection{Scattering on a step: formalism}
To solve the problem, we first obtain the modes on each side of the step. For this we define the transverse profiles $\vec{\Phi}^L_j$ ($j=1..2N_y-1$) as in section~\ref{Modes_Sec} with $N_y$ unit cells vertically, and $\vec{\Phi}^R_{j'}$ ($j'=1..2N_y-2N_0-1$) with $N_y-N_0$ unit cells vertically. Correspondingly, we define $\psi_{A/B,L}$ and $\psi_{A/B,R}$ the one-dimensional horizontal Bloch amplitudes on the left and right. We now consider a mode incoming from the left. This mode will be reflected, transmitted and converted into outgoing modes on both sides of the step. The scattering solution of this problem reads 
\bsub \label{2DWG_ScattSol} \bea
P_m &=& \sum_{j=1}^{2N_y-1}  a_j \lam_{+L,j}^m \bmat \psi_{A,L}^{+j} \vec{\Phi}^L_j \\ \psi_{B,L}^{+j} \vec{\Phi}^L_j \emat + b_j \lam_{-L,j}^m \bmat \psi_{A,L}^{-j} \vec{\Phi}^L_j \\ \psi_{B,L}^{+j} \vec{\Phi}^L_j \emat , \qquad (m<0) \label{2DWG_ScattSol_L} \\
&=& \sum_{j'=1}^{2N_y-2N_0-1}  c_{j'} \lam_{+R,j'}^m \bmat \psi_{A,R}^{+j'} \vec{\Phi}^R_{j'} \\ \psi_{B,R}^{+j'} \vec{\Phi}^R_{j'} \emat . \qquad (m \geqslant 0) 
\eea \esub
The $a_j$ are the incoming amplitudes, and the scattering matrices gives the others: $b_j = \sum_i [\hat{r}]_{ji} a_i$ and $c_j = \sum_i [\hat{t}]_{ji} a_i$. We now write matching conditions at $m=0$: for $n=N_0+1..N_y$ the pressure is ``continuous'' (i.e. still given by \eqref{2DWG_ScattSol_L}), and for $n=1..N_0$, the $\alpha$ and $\beta$ sites have zero pressure (open ends). This now gives the set of conditions 
\bsub \label{2DWG_step_Acond} \bea
\sum_{j=1}^{2N_y}  a_j \psi_{A,L}^{+j} \vec{\Phi}_j^L + b_j \psi_{A,L}^{-j} \vec{\Phi}_j^L &=& \sum_{j'=1}^{2N_y-2N_0}  c_{j'} \psi_{A,R}^{+j'} \vec{\Phi}_{j'}^R  \quad (n=N_0+1..N_y) , \\
\sum_{j=1}^{2N_y}  a_j \psi_{A,L}^{+j} \vec{\Phi}_j^L + b_j \psi_{A,L}^{-j} \vec{\Phi}_j^L &=& 0 \quad (n=1..N_0) ,
\eea \esub
and 
\be \label{2DWG_step_Bcond}
\sum_{j=1}^{2N_y}  a_j \psi_{B,L}^{+j} \vec{\Phi}_j^L + b_j \psi_{B,L}^{-j} \vec{\Phi}_j^L = \sum_{j'=1}^{2N_y-2N_0}  c_{j'} \psi_{B,R}^{+j'} \vec{\Phi}_{j'}^R  \quad (n=N_0+1..N_y) . 
\ee
It is important to notice the disymmetry of the conditions \eqref{2DWG_step_Acond} and \eqref{2DWG_step_Bcond}: we have no information about the value of the $B$-part pressure field ($\gamma$ and $\delta$ sites) on the ghost sites at $m=0$ and $n=1..N_0$. This prevents us from using the closure relation 
\be \label{2DWG_closureL}
\sum_{j=1}^{2N_y-1} | \vec{\Phi}_j^L \rangle \langle \vec{\Phi}_j^L | = \hat I_{2N_y}, 
\ee
on the $B$-condition \eqref{2DWG_step_Bcond}. One can use it on the $A$-condition \eqref{2DWG_step_Acond} however, as it corresponds to adding $0$'s to the missing components of $\vec{\Phi}_j^L$. On the other hand, the closure relation 
\be \label{2DWG_closureR}
\sum_{j=1}^{2N_y-2N_0-1} | \vec{\Phi}_j^R \rangle \langle \vec{\Phi}_j^R | = \hat I_{2N_y-2N_0}, 
\ee
can be used on both conditions. This disymmetry between the left and right side of the step is a standard difficulty when dealing with mode matching for cross-section changes in waveguides~\cite{Pagneux96}. Now, to obtain the scattering matrices, we apply the left-closure \eqref{2DWG_closureL} to the $A$-condition \eqref{2DWG_step_Acond} and the right-closure \eqref{2DWG_closureR} to the $B$-condition \eqref{2DWG_step_Bcond}. This gives us 
\be \label{2DWG_step_Acond_proj}
a_j \psi_{A,L}^{+j} + \left(\sum_i [\hat{r}]_{ji} a_i\right) \psi_{A,L}^{-j} = \sum_{j'=1}^{2N_y-2N_0-1} \left(\sum_i [\hat{t}]_{j'i} a_i\right) \psi_{A,R}^{+j'} \langle \vec{\Phi}_j^L | \vec{\Phi}_{j'}^R \rangle \quad (j=1..N_y), 
\ee
and
\be \label{2DWG_step_Bcond_proj}
\sum_{j=1}^{2N_y-1} a_j \psi_{B,L}^{+j} \langle \vec{\Phi}_{j'}^R | \vec{\Phi}_{j}^L \rangle + \left(\sum_i \hat{r}_{ji} a_i\right) \psi_{B,L}^{-j} \langle \vec{\Phi}_{j'}^R | \vec{\Phi}_{j}^L \rangle = \left(\sum_i [\hat{t}]_{j'i} a_i\right) \psi_{B,R}^{+j'}. 
\ee
We now define the $(2N_y-2N_0-1) \times (2N_y-1)$ matrix $Q$: 
\be
Q_{j'j} = \langle \vec{\Phi}_{j'}^R | \mathcal{P} \cdot \vec{\Phi}_{j}^L \rangle,  
\ee
where $\mathcal{P}$ is the operator projecting on the $N_0+1..N_y$ components. The disymmetry problem mentioned above is materialized here by the fact that $Q$ is not invertible. More precisely, $Q \cdot Q^\dagger = \hat I_{2N_y-2N_0}$ but $Q^\dagger \cdot Q \neq \hat I_{2N_y}$. We also define the normalization diagonal matrices 
\be \label{2DWG_Step_NormMat}
D_{A/B,L}^{\pm} = \mathrm{diag} \left(\psi_{A/B,L}^{\pm j}\right)_{j=1..2N_y-1} \quad \text{and} \quad D_{A/B,R}^{\pm} = \mathrm{diag}\left(\psi_{A/B,L}^{\pm j'}\right)_{j'=1..2N_y-2N_0-1}. 
\ee
The two matching conditions \eqref{2DWG_step_Acond_proj} and \eqref{2DWG_step_Acond_proj} gives us 
\bsub \bea
D_{A,L}^+ + D_{A,L}^- \cdot \hat{r} &=& Q^\dagger \cdot D_{A,R}^+ \cdot \hat{t} , \\
Q \cdot D_{B,L}^+ + Q \cdot D_{B,L}^- \cdot \hat{r} &=& D_{B,R}^+ \cdot \hat{t} .
\eea \esub
Solving this system leads to the scattering matrices: 
\bsub \label{Smat_inL} \bea
\hat{t} &=& (D_{B,R}^+ - Q \cdot D_{B,L}^- \cdot [D_{A,L}^-]^{-1} \cdot Q^\dagger \cdot D_{A,R}^+)^{-1} \cdot Q  \nonumber \\
&& \cdot (D_{B,L}^+ - D_{B,L}^-\cdot [D_{A,L}^-]^{-1} \cdot D_{A,L}^+ ), \\
\hat{r} &=& [D_{A,L}^-]^{-1} \cdot Q^\dagger \cdot D_{A,R}^+ \cdot \hat{t}  - [D_{A,L}^-]^{-1} \cdot D_{A,L}^+. 
\eea \esub
Following a similar method, we compute the scattering coefficients of waves coming in from the other side. Denoting them $\tilde{t}$ and $\tilde{r}$, they read 
\bsub \label{Smat_inR} \bea
\tilde{t} &=& (D_{A,L}^- - Q^\dagger \cdot D_{A,R}^+ \cdot [D_{B,R}^+]^{-1} \cdot Q \cdot D_{B,L}^-)^{-1} \cdot Q^\dagger  \nonumber \\
&& \cdot (D_{A,R}^- - D_{A,R}^+ \cdot [D_{B,R}^+]^{-1} \cdot D_{B,R}^- ), \\
\tilde{r} &=& [D_{B,R}^+]^{-1} \cdot Q \cdot D_{B,L}^- \cdot \tilde{t}  - [D_{B,R}^+]^{-1} \cdot D_{B,R}^-. 
\eea \esub
Notice that the reciprocity of the system imposes that after projection on the propagating modes subspaces, one has $\tilde{t} = \hat{t}^{\; T}$~\cite{Pagneux04}. Notice that to obtain the scattering matrices of a step where the right side is wider (formally $N_0<0$), one can use the above relation after the replacements $L \leftrightarrow R$, $+ \leftrightarrow -$ and $(\hat{r}, \hat{t}) \leftrightarrow (\tilde{r}, \tilde{t})$, but also by shifting the horizontal origin: $m\to m-1$. This is because the boundary condition along the vertical edge (using ghost cells) applies for $m=-1$ in this case. 

As we discussed at the beginning of this section, in the energy range $|t-s| < \varep < 1$, two of the eigenmodes are edge waves: one traveling to the left and the other to the right. Moreover, if we further assume that the energy lies inside the gap $1-|t-s| < \varep < 2|t-s|$, as in Fig.~\ref{WG_DispRel_Fig}-(a), all other modes are evanescent. Hence, an incoming edge wave from the left can only be reflected into a left moving edge wave with coefficient $R$ and transmitted into a right moving edge wave on the other side of the step with a coefficient $T$. Assuming the edge wave is the mode with index $j_d \in \mathbb Z$, we have $R=\hat{r}_{j_d j_d}$ and $T = \hat{t}_{j_d j_d}$. Moreover, as we explain in appendix~\ref{Current_App}, there is an energy current that is conserved through the scattering process. This implies conservation laws on the scattering coefficients. In the above energy range, where only edge waves propagate, this leads to the commun relation: 
\be \label{Scatt_Econs}
|T|^2 + |R|^2 = 1. 
\ee
Notice that since the system is reciprocal and energy conserving, the modulus of transmission and reflection is independent of whether the incident wave comes from the left or right side.

%%%%%%%%%%%%%%%%%%%%%%%%%%%%%%%%%%%%%%%%%%%%%%%%%%%
%						2D SCATTERING ON A STEP
%%%%%%%%%%%%%%%%%%%%%%%%%%%%%%%%%%%%%%%%%%%%%%%%%%%
\subsubsection{Scattering on a step: results}

\begin{figure}[htp]
\centering
\includegraphics[width=0.45\columnwidth]{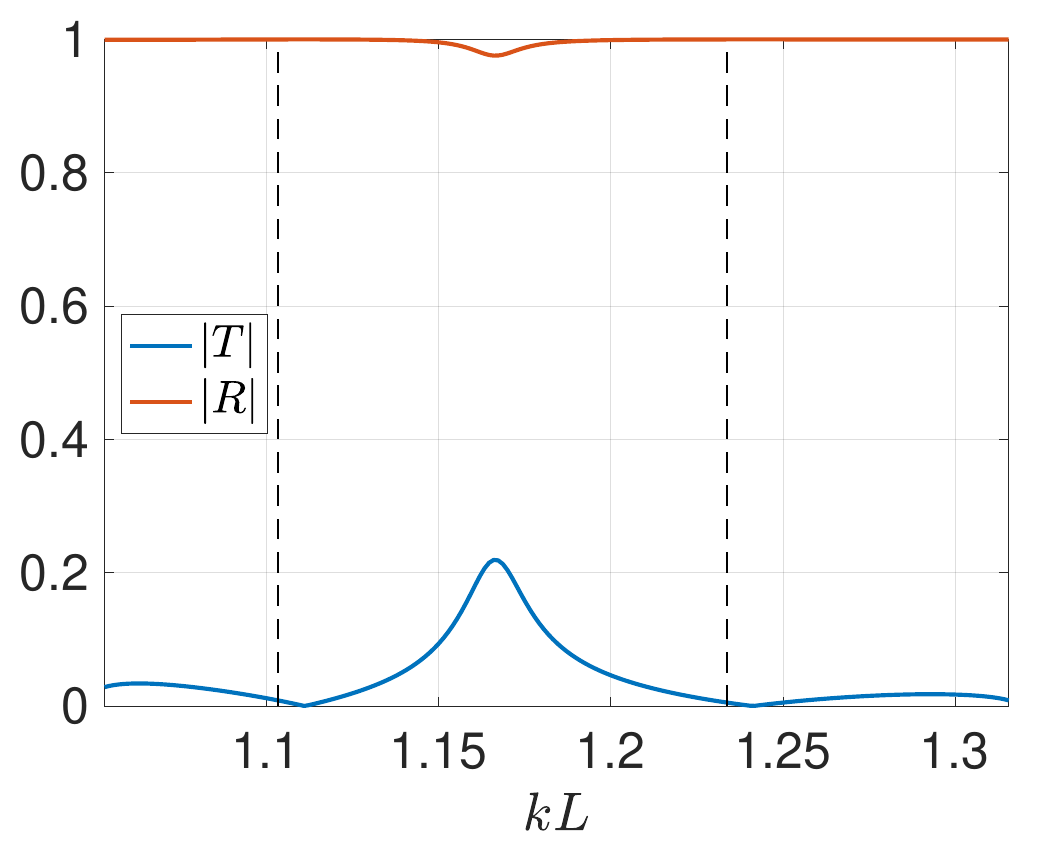} 
\includegraphics[width=0.45\columnwidth]{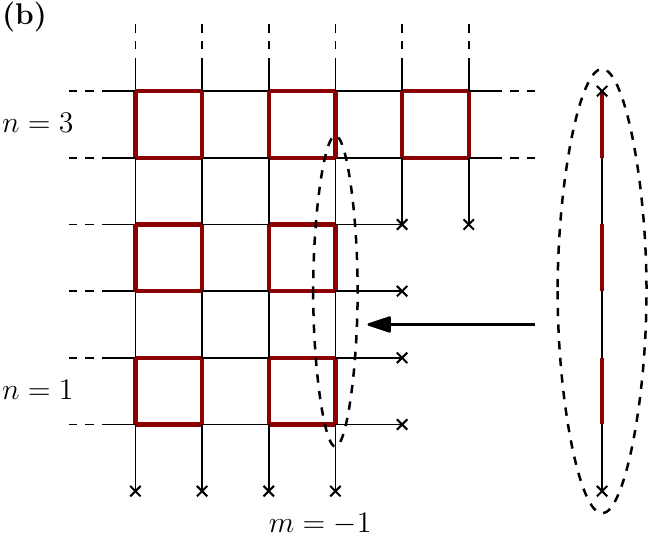} 
\caption{(a) Scattering coefficients $T$ and $R$ of a down wave on a step with $s=0.25$, $t=0.75$, $N_y=10$, and $N_0=2$. Dashed vertical lines show the eigenvalues of the 1D edge of the step. (b) Schematic of the scattering configuration, with emphasize on the effective 1D problem on the edge.}
\label{WG_ScattMat_Fig} 
\end{figure}

\begin{figure}[htp]
\centering
\includegraphics[width=\columnwidth]{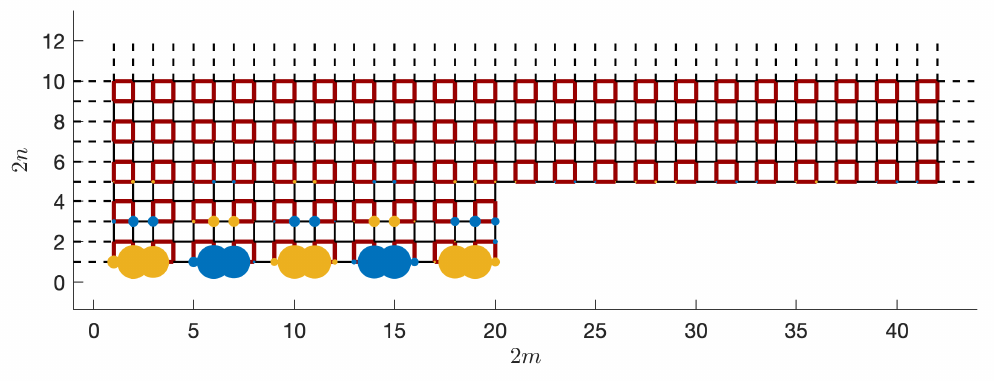}
\caption{Pressure amplitude $\Re(P_m)$ (arbitrary units) at each intersection of the network for a scattering on a step with $s=0.25$, $t=0.75$, $N_y=10$ and $N_0=2$. To increase visibility we only show the first $5$ unit cells vertically. Notice that although the reduced frequency $kL = 1.16$ is near the maximum transmission (see Fig.~\ref{WG_ScattMat_Fig}), the pressure amplitude is barely visible on the other side of the step.}
\label{WG_ScattMode_Fig} 
\end{figure}

The results for the scattering of an edge wave on a step of $N_0=2$ are shown in Fig.~\ref{WG_ScattMat_Fig}. We focus on the case where the full energy range of the edge wave is inside the gap, as in Fig.~\ref{WG_DispRel_Fig}-(a), that is, we assume both $1-|t-s| < E < 2|t-s|$ and $|t-s| < E < 1$. If we were considering an energy such that bulk waves coexist with the edge wave (which can happen for Fig.~\ref{WG_DispRel_Fig}-(b) and (c)), then the step would induce scattering in the bulk. The first remarkable fact in Fig.~\ref{WG_ScattMat_Fig}-(a) is that the edge wave is nearly perfectly reflected on the step, despite its rather small size. This means that edge waves on the lower edge are strongly decoupled from edge waves on the vertical edge of the step. In other words, in a 2D SSH network, edge waves are robustly confined to their natural boundary. We also show the profile of the scattering solution in Fig.~\ref{WG_ScattMode_Fig}. 

In addition, we see that transmission coefficient oscillates between zero and (small) maxima, indicating the presence of resonances. By comparing with the spectrum of a one-dimensional problem associated with the vertical edge of the step (see Fig.~\ref{WG_ScattMat_Fig}-(b)), we see that when the incident frequency is close to an eigenfrequency of the edge of the step, the transmission vanishes. This strengthen the conclusion that edge waves of vertical and horizontal edge are strongly decoupled. 

\begin{figure}[htp]
\centering
\includegraphics[width=\columnwidth]{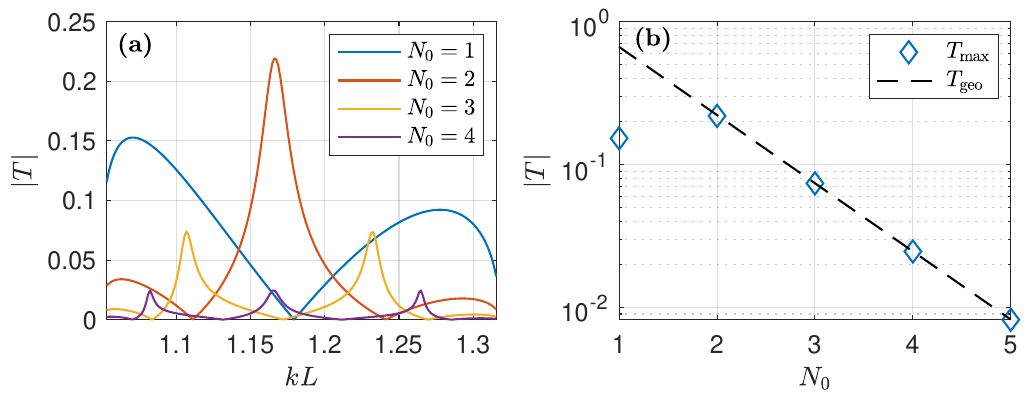} 
\caption{(a) Transmission coefficients on steps of various heights with $s=0.25$, $t=0.75$ and $N_y=10$. (b) Maximum of transmission $\max(|T|)$ (blue diamonds) of the edge wave in the gap ($|t-s| < \varep < 1$), with $s=0.25$ and $t=0.75$. The dashed black line is the guess $T_{\rm geom} = 2(s/t)^{N_0}$.}
\label{WG_ExpScale_Fig} 
\end{figure}

We also plot the transmission for several step heights in Fig.~\ref{WG_ExpScale_Fig}-(a). We see that for all step heights, transmission of edge waves stays low, which means that reflection is nearly perfect. In Fig.~\ref{WG_ExpScale_Fig}-(b), we look at the scaling of the maximum transmission as a function of the step size $N_0$. The latter decreases exponentially with $N_0$, which suggests that the non-zero transmission is achieved by evanescent coupling (tunnel effet). To confirm this, we compare the obtained result with an educated guess of $T_{\rm geo} = 2(s/t)^{N_0}$. To obtain this estimate, we use the fact that the penetration length of an edge wave is $1/\ln(s/t)$ in unit cell number, as we see in \eq{2D_sinf_EW}. Hence, the decrease of amplitude at the top of the step is $\propto (s/t)^{N_0}$, and we anticipate a transmission coefficient proportional to that decrease. Adding a (phenomenological) factor 2 to define $T_{\rm geo}$ leads to a rather good agreement.

\begin{figure}[htp]
\centering
\includegraphics[width=0.49\columnwidth]{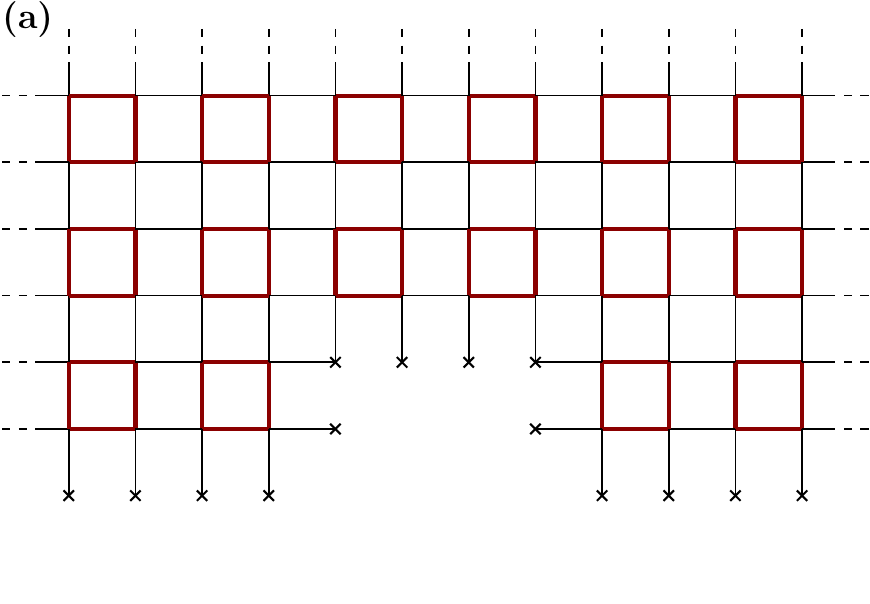} 
\includegraphics[width=0.49\columnwidth]{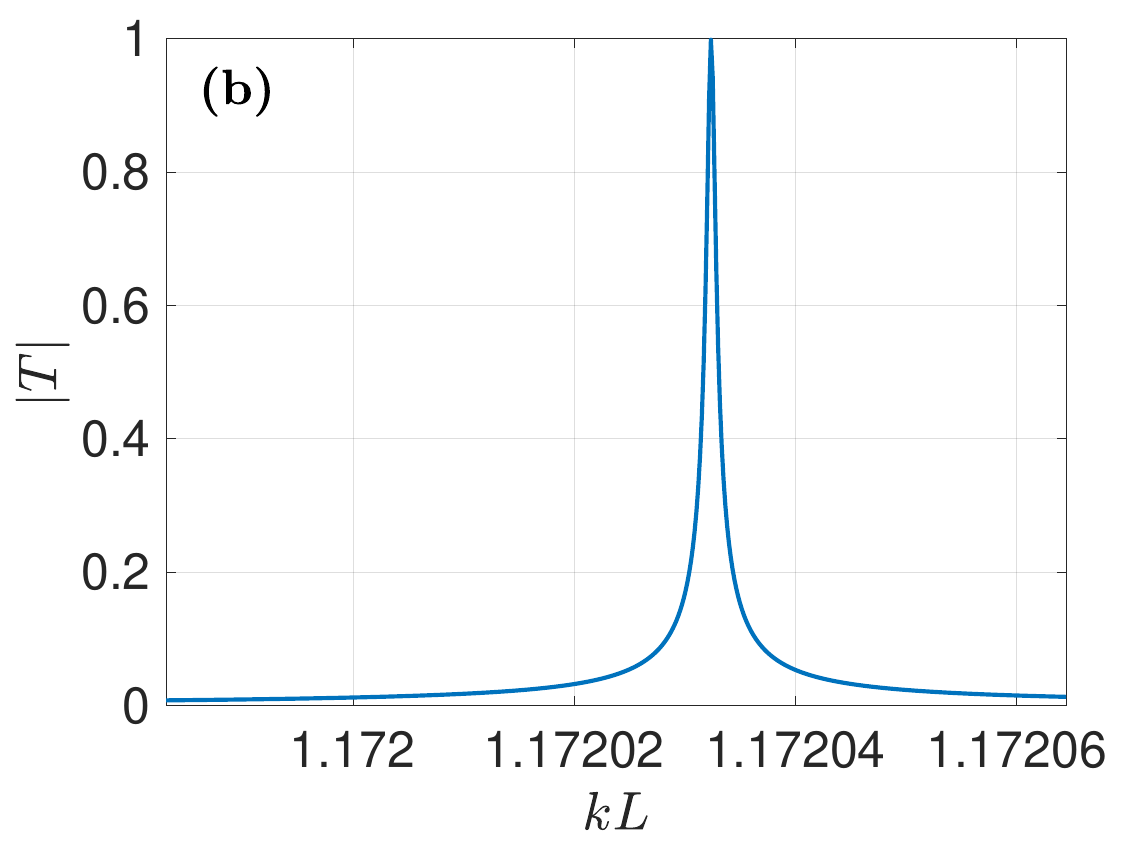}
\caption{(a) Transmission coefficients on a defect of height $N_0=1$ and length $N_d = 2$ unit cells in a network with $s=0.25$, $t=0.75$ and $N_y=10$. (b) Network with the defect of (a).}
\label{WG_DefectScatt_Fig} 
\end{figure}

Next, we consider the scattering of an edge wave on a defect consisting in a rectangle made of a few missing unit cells along the lower boundary, as shown in Fig.~\ref{WG_DefectScatt_Fig}-(a). This amounts to a combination of two steps: one of size $N_0$ and the other of size $-N_0$ separated by $N_d$ units cells. The scattering coefficients are obtained by using the preceding formulas \eqref{Smat_inL} and \eqref{Smat_inR} and the $S$-matrix product to combine them~\cite{Soukoulis}. The resulting transmission coefficient is shown in Fig.~\ref{WG_DefectScatt_Fig}-(b). We see that transmission is very low except in the close vicinity of a resonance frequency. This correspond to a resonance with the cavity induced by the defect. The resonance leads to perfect transmission, but only in a very narrow range of frequency. This explain for instance what was observed in~\cite{Liu18}. Out-of-resonance, transmission stays very low, in agreement with our previous results. Of course, since the resonance peak is very narrow, adding a small amount of dissipation will reduce it drastically, and the transmission will be very low across the whole range of frequency.

%%%%%%%%%%%%%%%%%%%%%%%%%%%%%%%%%%%%%%%%%%%%%%%%%%%
%%%%%%%%%%%%%%%%%%%%%%%%%%%%%%%%%%%%%%%%%%%%%%%%%%%
%%%%%%%%%%%%%%%%%%%%%%%%%%%%%%%%%%%%%%%%%%%%%%%%%%%
%
%						2D finite network
%
%%%%%%%%%%%%%%%%%%%%%%%%%%%%%%%%%%%%%%%%%%%%%%%%%%%
%%%%%%%%%%%%%%%%%%%%%%%%%%%%%%%%%%%%%%%%%%%%%%%%%%%
%%%%%%%%%%%%%%%%%%%%%%%%%%%%%%%%%%%%%%%%%%%%%%%%%%%
\section{Consequences for modes in finite networks}
\label{Finite_Net_Sec}
\begin{figure}[htp]
\centering
\includegraphics[width=0.5\columnwidth]{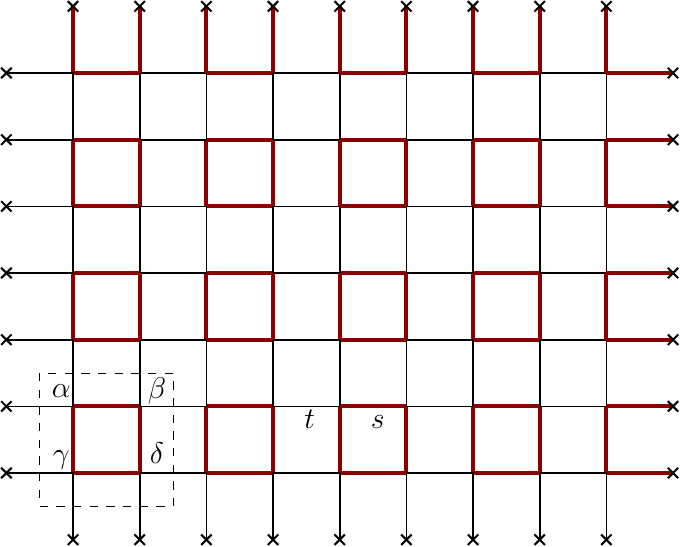}
\caption{Finite two-dimensional SSH networks for $N_x=4$ and $N_y=3$. }
\label{2D_SSH_Finite_Fig} 
\end{figure}

We now investigate the consequences of the perfect reflection of edge waves at corners in a finite network of $N_x$ cells in the $x$ direction and $N_y$ cells in the $y$ direction, as shown in Fig.~\ref{2D_SSH_Finite_Fig}. the lower and upper edge are obtained as before (see beginning of section~\ref{Waveguide_Sec}), and the left (resp. right) edge is obtained as the lower (resp. upper) one. We underline that in this construction, for $s<t$, topological edge waves can appear only in the lower and left edges.

\subsection{Rectangular networks}

\begin{figure}[htp]
\centering
\includegraphics[width=0.45\columnwidth]{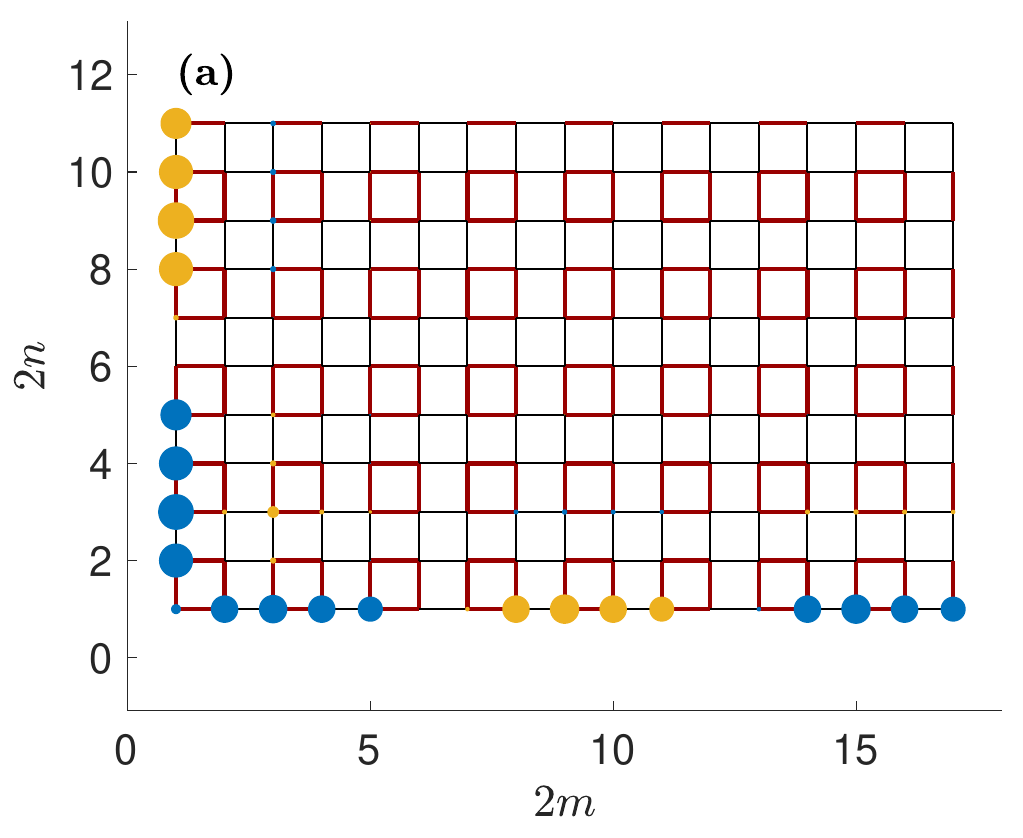}
\includegraphics[width=0.45\columnwidth]{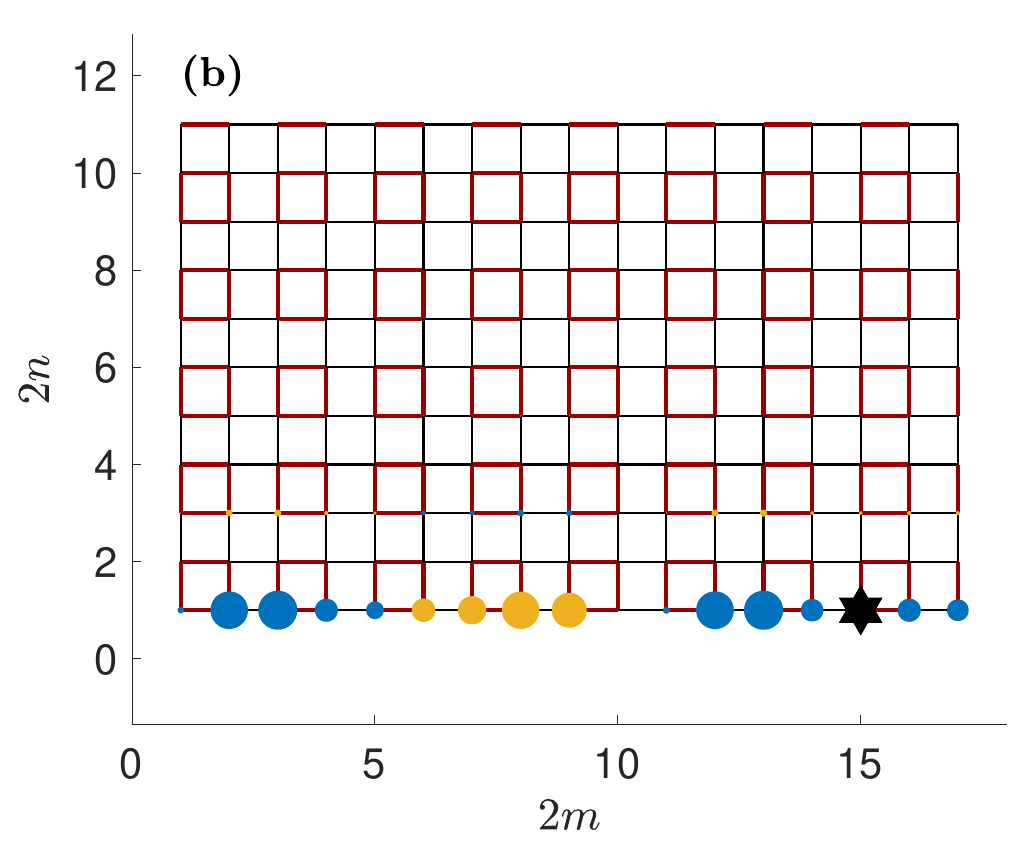} 
\caption{Pressure amplitudes (arbitrary units) in a network with $N_x=8$, $N_y=5$, $s=0.15$, and $t=0.85$. Positive (resp. negative) values of the amplitude are shown in yellow (resp. blue). (a) eigenmode with $kL=1.085$. (b) Field generated by a pressure source (star) with $kL=1.1$ (value chosen to be close to that of (a)).}
\label{Finite_EigModes_Fig} 
\end{figure}

When looking at the eigenmodes of a finite network, we expect from the preceding sections that edge modes will be localized either on the lower edge or on the left edge, since a corner is perfectly reflecting. However, if eigenmodes localized on the lower and left edges share the same eigenfrequency (degeneracy), we will witness eigenmodes localized on both edges. This would give a false impression that transmission between edges is possible, as shown in Fig.~\ref{Finite_EigModes_Fig}-(a). In the 2D SSH model, degeneracy happens rather often. Let us first discuss the case of a square network $N_x=N_y$. The problem is symmetric with respect to the diagonal of the square network, which means that each edge wave energy level is doubly degenerate, with one mode being localized on the lower edge and the other on the left edge. In this case, every linear combination of the two is an equally valid eigenmode, and is localized on both edges. In a rectangular network, degenerate energy levels can arise for similar reasons. In~\cite{Coutant20}, we have shown that if $N_d$ is the greatest common divisor (gcd) of $N_x+1$ and $N_y+1$, there are $N_d$ pairs of degenerate edge waves (see section~II-B of~\cite{Coutant20}). This is the case for the network of Fig.~\ref{Finite_EigModes_Fig} with $N_d=\mathrm{gcd}(9,6)=3$. 

On the contrary, one can look at the response of a point source localized at a given network intersection with a reduced frequency close to an eigenfrequency of an edge wave. This is done by writing the eigenvalue problem \eqref{2D_SSH_master_Eq} as $\varep P = H \cdot P$, where $H$ is a square hermitian matrix. Considering a point source $S$ localized on a specific site, the response is given by the Green function: 
\be
P = (\varep(kL) - H)^{-1} \cdot S . 
\ee
When the reduced frequency is in the gap and the edge wave range, the obtained response field will be localized only on the closest edge to the source, which confirms the absence of transmission from one to the other. This is shown in Fig.~\ref{Finite_EigModes_Fig}-(b). 

\subsection{Networks with defects}

\begin{figure}[htp]
\centering
\includegraphics[width=0.49\columnwidth]{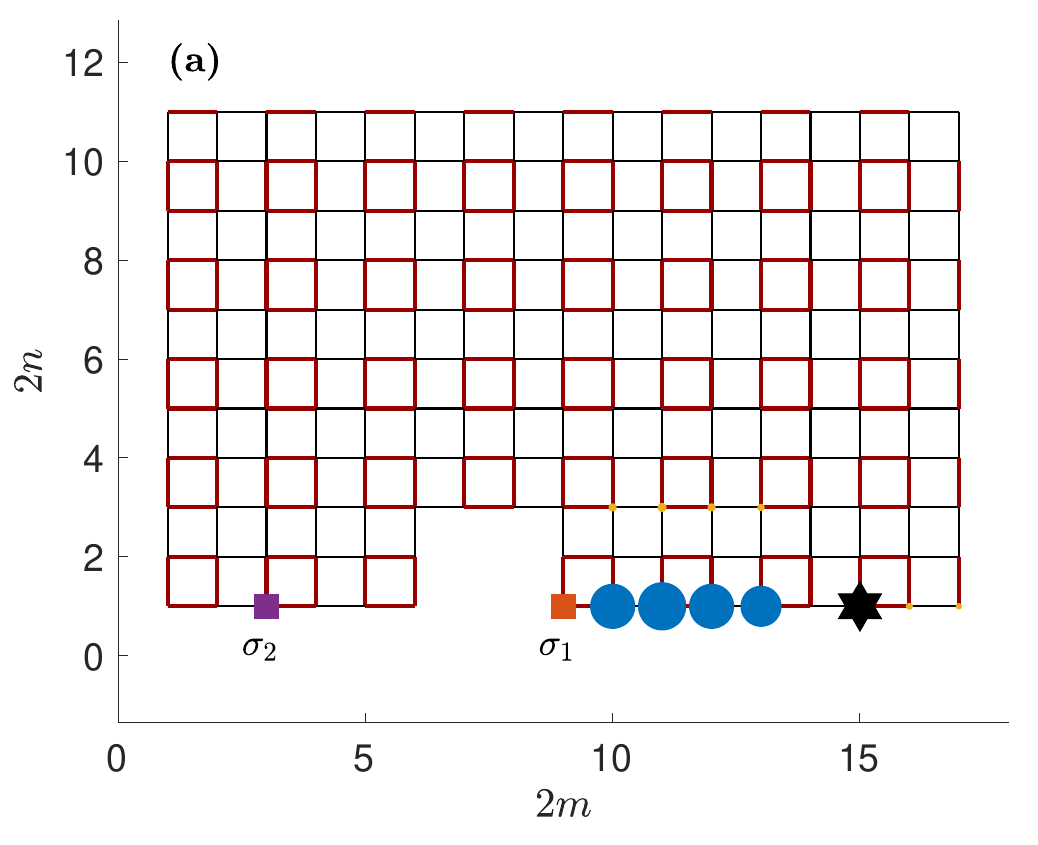} 
\includegraphics[width=0.49\columnwidth]{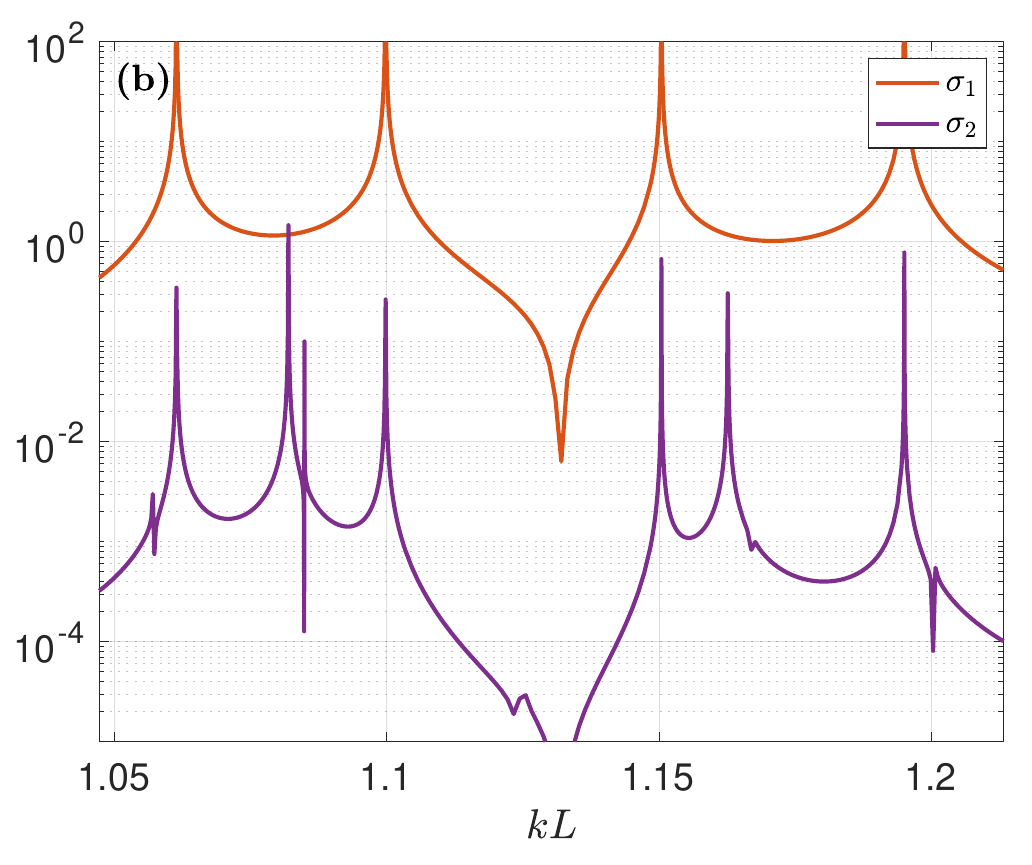}
\caption{Pressure amplitudes (arbitrary units) from a source located at the black star in a network with $N_x=8$, $N_y=5$, $s=0.25$, and $t=0.75$. (a) $kL=1.085$. Positive (resp. negative) values of the amplitude are shown in yellow (resp. blue). (b) Pressure at two intersections $\sigma_1$ and $\sigma_2$ marked by squares in (a) and for varying reduced frequency $kL$.}
\label{Finite_EigModes_Defect_Fig} 
\end{figure}

\begin{figure}[htp]
\centering
\includegraphics[width=0.49\columnwidth]{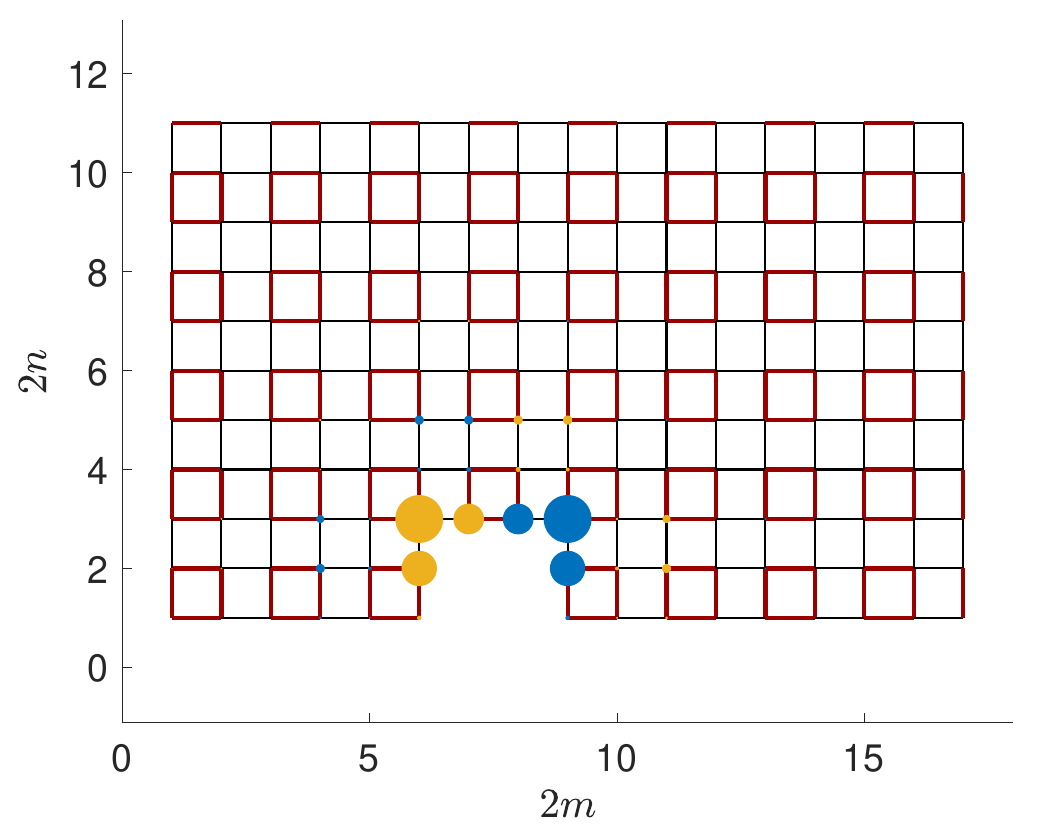} 
\caption{Pressure amplitudes (arbitrary units) of an eigenmode localized on the edge on top of a missing cell. We took a network with $N_x=8$, $N_y=5$, $s=0.25$, and $t=0.75$. The associated eigenfrequency is $kL=0.956$.}
\label{Finite_DefectMode_Fig} 
\end{figure}

We now consider a finite network with a defect consisting in missing unit cells along the lower edge. We put a source on the lower right side, and a receptor on each side of the defect: $\sigma_1$ on the same side as the source and $\sigma_2$ on the other side. This is shown in Fig.~\ref{Finite_EigModes_Defect_Fig}. When scanning in energy $\varep$, we see peaks of transmission to $\sigma_1$ near each edge wave eigen-energy. On the contrary, we see essentially no transmission to the receptor $\sigma_2$ on the other side, which is again explained by the quasi perfect reflection on the defect. However, we see in Fig.~\ref{Finite_EigModes_Defect_Fig}-(b) extremely narrow resonances where transmission to $\sigma_2$ do occur. These resonances correspond to edge eigenmodes of the effective edges on each side of the defect (which in turn are good approximation of the exact eigenvalues since the two are essentially decoupled). However, the resonances allowing for transmission to $\sigma_2$ are so narrow that a tiny amount of dissipation in the system will suppress them. As a last remark, we point out that since every straight part of the edges act as an isolated cavity for edge waves, due to high relfectivity, there are also edge waves hosted on the top of the missing cell. This is shown in Fig.~\ref{Finite_DefectMode_Fig}. 

%%%%%%%%%%%%%%%%%%%%%%%%%%%%%%%%%%%%%%%%%%%%%%%%%%%
%%%%%%%%%%%%%%%%%%%%%%%%%%%%%%%%%%%%%%%%%%%%%%%%%%%
%%%%%%%%%%%%%%%%%%%%%%%%%%%%%%%%%%%%%%%%%%%%%%%%%%%
%
%							CONCLUSION
%
%%%%%%%%%%%%%%%%%%%%%%%%%%%%%%%%%%%%%%%%%%%%%%%%%%%
%%%%%%%%%%%%%%%%%%%%%%%%%%%%%%%%%%%%%%%%%%%%%%%%%%%
%%%%%%%%%%%%%%%%%%%%%%%%%%%%%%%%%%%%%%%%%%%%%%%%%%%
\section{Conclusion}

In this work, we analyze the scattering properties of topological edge waves in an acoustic realization of the two-dimensional SSH model~\cite{Liu17,Liu18,Obana19,Zheng19}. We first analytically show that an incident edge wave on an isolated corner exactly undergoes a total reflection (see \eq{Corner_Scatt_eq}). Moreover, the phase of the reflection coefficient can be used as a topological marker, to attest for the presence of corner modes (i.e. second order topological insulator phase), see Fig.~\ref{Corner_Phase_Fig}. We then study the scattering of edge waves on more general changes of edge structure: steps and defects. We find that edge waves are largely reflected on all types of changes as shown in Figs.~\ref{WG_ScattMat_Fig} and \ref{WG_DefectScatt_Fig}. Although reflection is no longer total, it approaches unity exponentially fast as the change size is increased. This is explained by the fact that transmission across steps of defect is only possible through evanescent coupling. We then study consequences of this strong reflections for finite networks. We show that each turn or defect splits an edge into two decoupled straight edges that can host independent sets of edge waves, as shown in Figs.~\ref{Finite_EigModes_Fig} and \ref{Finite_EigModes_Defect_Fig}. This work demonstrates that topological edge waves of the 2D SSH model are immune to forward scattering on corners or defect. This provides the interesting possibility for isolation of edge waves in specific parts of the boundary of a network.

\begin{acknowledgments}
The data that support the findings of this study are available from the corresponding author upon reasonable request. 
This project has received funding from the European Union's Horizon 2020 research and innovation programme under the Marie Sklodowska-Curie grant agreement No 843152. 
\end{acknowledgments}

%%%%%%%%%%%%%%%%%%%%%%%%%%%%%%%%%%%%%%%%%%%%%%%%%%%
%%%%%%%%%%%%%%%%%%%%%%%%%%%%%%%%%%%%%%%%%%%%%%%%%%%
%%%%%%%%%%%%%%%%%%%%%%%%%%%%%%%%%%%%%%%%%%%%%%%%%%%
%
%							APPENDIX
%
%%%%%%%%%%%%%%%%%%%%%%%%%%%%%%%%%%%%%%%%%%%%%%%%%%%
%%%%%%%%%%%%%%%%%%%%%%%%%%%%%%%%%%%%%%%%%%%%%%%%%%%
%%%%%%%%%%%%%%%%%%%%%%%%%%%%%%%%%%%%%%%%%%%%%%%%%%%
\newpage
\appendix

%%%%%%%%%%%%%%%%%%%%%%%%%%%%%%%%%%%%%%%%%%%%%%%%%%%
%						            Separability
%%%%%%%%%%%%%%%%%%%%%%%%%%%%%%%%%%%%%%%%%%%%%%%%%%%
\section{Separability of the 2D SSH model}
\label{Sep_App}

In this appendix, we detail how the 2D SSH model can be solved by separation of variables, see also \cite{Zhu20,Coutant20} for an approach with the Kronecker tensor product. As stated at the beginning of section~\ref{Bloch_Sec}, solution of equations~\eqref{2D_SSH_master_Eq} can be found of the form 
\be \label{App_Sep_eq}
\bmat \pal_{m,n} \\ \pbet_{m,n} \\ \pgam_{m,n} \\ \pdel_{m,n} \emat = \bmat \psi_A^m \varphi_A^n \\ \psi_A^m \varphi_B^n \\ \psi_B^m \varphi_A^n \\ \psi_B^m \varphi_B^n \emat . 
\ee 
This is nothing else than the discrete equivalent of looking for solutions of the Helmholtz equation of the form $f(x,y) = g(x) h(y)$. In fact, the intracell indices makes this statement slightly less transparent, but we can relabel every amplitudes by vertical and horizontal integer $\bar m$ and $\bar n$ to make the analogy with the Helmholtz equation straightforward. For this, we define $\alpha_{m,n} = f_{2\bar m-1, 2\bar n-1}$, $\beta_{m,n} = f_{2\bar m-1, 2\bar n}$, $\gamma_{m,n} = f_{2\bar m, 2\bar n-1}$, and $\delta_{m,n} = f_{2\bar m, 2\bar n}$, then the separation of variable assumption of \eq{Full_2D_TensorProd} or \eqref{App_Sep_eq} simply reads $f_{\bar m, \bar n} = g_{\bar m} h_{\bar n}$. We can then use this separation of variable assumption in the main equations~\eqref{2D_SSH_master_Eq}. We do this explicitly for equation~\eqref{2DSSH_alpha}: 
\bsub \bea
\varep \pal_{m,n} &=& s \pbet_{m,n} + t \pbet_{m,n-1} + s \pgam_{m,n} + t \pgam_{m-1,n}, \\
\varep \psi_A^{m} \varphi_A^{n} &=& s \psi_A^{m} \varphi_B^{n} + t \psi_A^{m} \varphi_B^{n-1} + s \psi_B^{m} \varphi_A^{n} + t \psi_B^{m-1} \varphi_A^{n}, \\
\varep \psi_A^{m} \varphi_A^{n} &=& \left( s \varphi_B^{n} + t \varphi_B^{n-1} \right) \psi_A^{m} + \left(s \psi_B^{m} + t \psi_B^{m-1} \right) \varphi_A^{n}, 
\eea \esub
hence, 
\be
\varep = \frac1{\varphi_A^{n}} \left( s \varphi_B^{n} + t \varphi_B^{n-1} \right) + \frac1{\psi_A^m} \left(s \psi_B^{m} + t \psi_B^{m-1} \right). 
\ee
In this last equation, the left-hand side is a constant independent of $m$ and $n$. Since the first term on the right-hand side only depend on $n$, and the second term only on $m$, both terms must be constant. By calling the first constant $\varep_y$ and the second $\varep_x$, we obtain 
\be
\varep_y \varphi_A^n = t \varphi_B^{n-1} + s \varphi_B^{n} , 
\ee
and
\be
\varep_x \psi_A^m = t \psi_B^{m-1} + s \psi_B^{m} .   
\ee 
This is half of the four one-dimensional equations~\eqref{Long_SSH} and \eqref{Transverse_SSH}. With a similar calculation for the three other equations in~~\eqref{2D_SSH_master_Eq}, we obtain the other two one-dimensional equations. Note that this is not trivial since the procedure leads to 8 equations from the 4 equations~\eqref{2D_SSH_master_Eq}, while there are only 4 one-dimensional equations. In particular, the 8 obtained equations reduce to 4 at the condition that the constants that appear are always the same two constants $E_x$ and $E_y$.

%%%%%%%%%%%%%%%%%%%%%%%%%%%%%%%%%%%%%%%%%%%%%%%%%%%
%						            Energy Current
%%%%%%%%%%%%%%%%%%%%%%%%%%%%%%%%%%%%%%%%%%%%%%%%%%%
\section{Energy current conservation}
\label{Current_App}
The model studied in this work possesses a conserved current, that we refer to as ``energy current''. The existence of this current is a consequence of the hermitian character of the Hamiltonian and the discrete translation invariance (see e.g.~\cite{Dwivedi16} for a discussion in lattice models). In the core of this work, we have normalized waveguide modes (see \eq{2DWG_mode}) so that they transport a unit amount of current. This is very convenient as it automatically gives relations among the scattering coefficients, such as \eq{Scatt_Econs}. 

To obtain a conserved current, we first use the equations~\eqref{2D_SSH_master_Eq} to compute $\varep (|\pal_{m,n}|^2+|\pbet_{m,n}|^2+|\pgam_{m,n}|^2+|\pdel_{m,n}|^2)$. This gives a sum of terms, half with intracell coupling $s$ as a commun factor and the other half with the intercell coupling $t$ as commun factor. Now we notice that the $s$ terms come in pairs of complex conjugate. Hence, since $\varep (|\pal_{m,n}|^2+|\pbet_{m,n}|^2+|\pgam_{m,n}|^2+|\pdel_{m,n}|^2)$ is manifestly real, the imaginary part of the sum of $t$ terms is zero. This gives us the equation 
\be \label{Local_Current}
0 = J^{x}_{m+1,n} - J^{x}_{m,n} + J^{y}_{m,n+1} - J^{y}_{m,n}, 
\ee
with 
\bsub \bea 
J^{x}_{m,n} &=& t \Im \left( (\pgam_{m,n})^* \pal_{m+1,n} + (\pdel_{m,n})^* \pbet_{m+1,n} \right) , \\
J^{x}_{m,n} &=& t \Im \left( (\pbet_{m,n})^* \pal_{m,n+1} + (\pdel_{m,n})^* \pgam_{m,n+1} \right) . 
\eea \esub 
Equation~\eqref{Local_Current} is a discrete equivalent of a local conservation law of the form $\mathrm{div}(\mathbf{J}) = 0$. Now, in a waveguide configuration, we sum the local conservation \eqref{Local_Current} over $n$ from 0 to $N_y$. The terms in $J^{y}_{m,n}$ exactly add up to boundary terms $J^{y}_{m,n=N_y+1} - J^{y}_{m,n=0}$, which vanishes due to the boundary conditions. Hence we obtain the conserved waveguide current as: 
\be
J_m = -2t \sum_n \Im \left((\pgam_{m,n})^* \pal_{m+1,n} + (\pdel_{m,n})^* \pbet_{m+1,n}\right). 
\ee
This can be further simplifies by using again equations~\eqref{2D_SSH_master_Eq} (in particular \eqref{2DSSH_gamma} and \eqref{2DSSH_delta}) to get rid of amplitudes at $m+1$ in favor of amplitudes at $m$. This leads to 
\be
J_m = 2s \sum_n \Im \left((\pgam_{m,n})^* \pal_{m,n} + (\pdel_{m,n})^* \pbet_{m,n}\right). 
\ee
To explicitly compute the current on a particular waveguide mode of the form \eqref{2DWG_mode}, we first notice that since $\vec{\Phi}_j$ are eigen-vectors of an hermitian eigenvalue problem, they can be chosen orthonormal, that is 
\be \label{Transverse_Norm}
\langle \vec{\Phi}_i | \vec{\Phi}_j \rangle = \delta_{ij}, 
\ee
with the canonical product
\be \label{2DWG_Ortho}
\langle \vec{\Phi}' | \vec{\Phi} \rangle = \sum_{n=1}^{N_y} (\varphi'_{A,n})^* \varphi_{A,n} + (\varphi'_{B,n})^* \varphi_{B,n}. 
\ee
We can now compute the current associated to a given mode as in \eq{2DWG_mode}, and we obtain 
\bea
J_m &=& 2s \Im \left((\psi_B^j)^* \psi_A^j \right) \times \langle \vec{\Phi}_j | \vec{\Phi}_j \rangle, 
\eea
where the second factor is unity because the transverse modes are normalized according to \eq{Transverse_Norm}. There are now two possibilities depending on whether the mode is evanescent or propagative. In the former case, we see from \eq{1D_SSH_Bloch} that $\Im \left((\psi_B^j)^* \psi_A^j \right)=0$. This is expected since an evanescent wave alone does not transport energy. For propagative modes, we can use \eq{1D_SSH_Bloch} to simplify the expression, and we obtain 
\bea
J_m &=& 2s \Im \left((\psi_B^j)^* \psi_A^j \right) , \nonumber \\
&=& -\sin(q_j) \frac{2st}{\varep_x} |\psi_B^j|^2 , \nonumber \\
&=& v_g^j \left(|\psi_A^j|^2 + |\psi_B^j|^2 \right). 
\eea
This result can be interpreted as follows: a propagative mode transport an energy proportional to its amplitude $|\psi_A^j|^2 + |\psi_B^j|^2$ at a speed given by the group velocity $v_g^j$. In the core of the paper, waveguide modes are normalized such that they transport a unit current, and hence, scattering coefficients obey conservation laws of the form of \eq{Scatt_Econs}.

%%%%%%%%%%%%%%%%%%%%%%%%%%%%%%%%%%%%%%%%%%%%%%%%%%%
%						            Symmetries
%%%%%%%%%%%%%%%%%%%%%%%%%%%%%%%%%%%%%%%%%%%%%%%%%%%
\section{Chiral symmetries}
\label{Chiral_App}
Just like its one-dimensional counterpart, the 2D SSH model is chiral symmetric. This means that there is a chiral operator that acts inside each cell and such that 
\be
\Gam \cdot H \cdot \Gam = - H.  
\ee
The chiral operator $\Gam$ is defined by  
\be
\Gam \cdot \bmat \pal_{m,n} \\ \pbet_{m,n} \\ \pgam_{m,n} \\ \pdel_{m,n} \emat = \bmat 1 & 0 & 0 & 0 \\ 0 & -1 & 0 & 0 \\ 0 & 0 & -1 & 0 \\ 0 & 0 & 0 & 1 \emat \cdot \bmat \pal_{m,n} \\ \pbet_{m,n} \\ \pgam_{m,n} \\ \pdel_{m,n} \emat, 
\ee
The consequences of this symmetry are the same as in 1D: the spectrum is symmetric about 0: if $P$ is an eigenvector with eigenvalue $\varep$, then $\Gam \cdot P$ is an eigenvector with eigenvalue $-\varep$. Hence, the spectrum is symmetric under $\varep \to -\varep$. This applies in particular to edge waves. In Fig.~\ref{WG_EdgeModes_Fig}-(b), we see that there are two bands of edge waves, one with $\varep > 0$ and another with $\varep < 0$. One can be deduced from the other simply by applying the operation $\Gam$.

\bibliographystyle{utphys}
\bibliography{Bibli}

\end{document}